\documentclass[10pt,aps,pra,twocolumn,superscriptaddress,floatfix,nofootinbib]{revtex4-2}

\makeatletter
\def\@bibdataout@aps{%
 \immediate\write\@bibdataout{%
  @CONTROL{%
   apsrev41Control,author="08",editor="1",pages="0",title="0",year="1",eprint="1"%
  }%
 }%
 \if@filesw
  \immediate\write\@auxout{\string\citation{apsrev41Control}}%
 \fi
}%
\makeatother 

\usepackage[T1]{fontenc}
\usepackage{amsfonts,amsmath,mathrsfs,mathtools}
\usepackage{siunitx}
\usepackage{booktabs}
\usepackage{multirow}
\usepackage{algorithm}
\usepackage{algpseudocode}
\usepackage{lipsum} 
\usepackage{printlen}
\usepackage{soul}

\usepackage{CJKutf8}
\usepackage[utf8]{inputenc} 
\usepackage[T1]{fontenc}

\usepackage[breaklinks=true]{hyperref}
\usepackage[usenames,dvipsnames]{color}
\hypersetup{
  colorlinks   = true, 
  urlcolor     = blue, 
  linkcolor    = blue, 
  citecolor   = red 
}
\usepackage{cleveref}
\usepackage{xy}
\xyoption{matrix}
\xyoption{frame}
\xyoption{arrow}
\xyoption{arc}

\usepackage{ifpdf}
\ifpdf
\else
\PackageWarningNoLine{Qcircuit}{Qcircuit is loading in Postscript mode.  The Xy-pic options ps and dvips will be loaded.  If you wish to use other Postscript drivers for Xy-pic, you must modify the code in Qcircuit.tex}
\xyoption{ps}
\xyoption{dvips}
\fi

\entrymodifiers={!C\entrybox}

\newcommand{\bra}[1]{{\left\langle{#1}\right\vert}}
\newcommand{\ket}[1]{{\left\vert{#1}\right\rangle}}
\newcommand{\qw}[1][-1]{\ar @{-} [0,#1]}
\newcommand{\qwx}[1][-1]{\ar @{-} [#1,0]}
\newcommand{\control}{*!<0em,.025em>-=-<.2em>{\bullet}}

\newcommand{\ctrl}[1]{\control \qwx[#1] \qw}

\newcommand{\targ}{*+<.02em,.02em>{\xy ="i","i"-<.39em,0em>;"i"+<.39em,0em> **\dir{-}, "i"-<0em,.39em>;"i"+<0em,.39em> **\dir{-},"i"*\xycircle<.4em>{} \endxy} \qw}
\newcommand{\lstick}[1]{*!R!<.5em,0em>=<0em>{#1}}

\newcommand{\Qcircuit}{\xymatrix @*=<0em>}

\newcommand{\half}{\tfrac{1}{2}}

\newcommand{\op}[2]{\ket{#1}\!\bra{#2}}        

\newcommand{\phiext}{\phi_{\rm ext}}

\newcommand{\zp}{$0$-$\pi$}
\newcommand{\zps}{$0$-$\pi$\ }

\crefformat{equation}{Eq.~(#2#1#3)} 
\crefformat{section}{Sec.~#2#1#3} 
\Crefformat{equation}{Equation~(#2#1#3)}
\crefformat{figure}{Fig.~#2#1#3}
\Crefformat{figure}{Figure~#2#1#3}
\crefrangeformat{equation}{Eqs.~#3(#1)#4--#5(#2)#6}
\crefmultiformat{equation}{Eqs.~(#2#1#3)}{ and~(#2#1#3)}{, (#2#1#3)}{ and~(#2#1#3)}

\makeatother

\begin{document}

\hfuzz=150pt
\hbadness=10000

\begin{CJK*}{UTF8}{gbsn} 
\title{
Two-qubit gates for the soft \texorpdfstring{\zp}{zero pi} qubit
}
\author{Zhenxing Liu}
\email[]{liuzx@baqis.ac.cn}
\affiliation{Beijing Key Laboratory of Fault-Tolerant Quantum Computing, Beijing Academy of Quantum Information Sciences, Beijing 100193, China}
\address{Materials Science and Engineering, University of Colorado Boulder, Colorado 80309, USA}
\author{Eli Weissler}
\address{Electrical, Computer, and Energy Engineering, University of Colorado Boulder, Colorado 80309, USA}
\author{Joshua Combes}
\address{Electrical, Computer, and Energy Engineering, University of Colorado Boulder, Colorado 80309, USA}
\affiliation{School of Physics and School of Mathematics, The University of Melbourne, VIC 3010, Australia}
\date{\today}

\begin{abstract}
The \zps qubit promises longer lifetimes than the transmon, making it a strong candidate for a next-generation, lower-error qubit.
Gyenis et al. experimentally demonstrated an unprotected single-qubit gate in the soft 0-$\pi$ regime~\cite{gyenis2021}, but no corresponding two-qubit gate has yet been proposed.
We propose an unprotected CZ gate for capacitively coupled soft \zps qubits. The gate uses a direct transition between a computational state and a higher non-computational state. Assuming negligible device disorder and a phenomenological noise model, the simulated CZ gate achieves a fidelity of approximately $99.9\%$ at a gate time of approximately $160~\mathrm{ns}$.
\end{abstract}
\maketitle
\end{CJK*}

\section{Introduction}
The transmon and fluxonium are the workhorses of superconducting quantum computing. They combine relatively simple fabrication with high performance ~\cite{koch2007,manucharyan2009fluxonium,nguyen2019high,arute2019quantum}. Their coherence times, however, are limited by a trade-off between relaxation and dephasing. The transmon suppresses charge-noise dephasing but is susceptible to energy relaxation. Heavy fluxonium is comparatively robust against relaxation but more sensitive to flux-noise dephasing. The standard approach to overcome this issue is to use error correction. The relaxation/dephasing trade-off is generic to single-mode superconducting qubits, which cannot be strongly protected against both relaxation and dephasing~\cite{gyenis2021moving}.

Multimode circuits offer an alternative route beyond this limitation. They can realize \textit{protected qubits}~\cite{gyenis2021moving} that are resistant to both energy relaxation and dephasing, and thus can lower error rates. However, this protection comes at the cost of increased circuit and fabrication complexity. Replacing transmons and fluxoniums with such protected qubits could reduce the error-correction overhead for large-scale quantum computation~\cite{eisert2026mindgapsfraughtroad}.

Among protected-qubit proposals, the \zps qubit is one of the most widely studied~\cite{kitaev2006}. Since its introduction~\cite{kitaev2006,brooks2013}, studies have examined disorder~\cite{dempster2014}, coherence~\cite{groszkowski2018}, readout and control~\cite{di2019,gyenis2021}, and tunable circuit designs~\cite{shen2015,smith2020,egusquiza2021,guo2022,rajpoot2022tunable}. The first experimental realization operated in a partially protected ``soft'' regime~\cite{gyenis2021}, with exponential protection against relaxation but only first-order protection against dephasing. Full protection requires large inductances and small stray capacitances, which pose major fabrication challenges. Although recent advances have moved experimental parameters closer to the ideal regime~\cite{KimAPS2024,kim2025protected}, near-term \zps devices are likely to remain soft, with steadily increasing protection.

Long-lived qubits alone are not enough for quantum computation. Universal control also requires high-fidelity single- and two-qubit gates. Fully protected gates are possible~\cite{kolesnikow2025} but are difficult to realize experimentally. Near-term \zps devices will likely rely on unprotected gates that do not preserve the qubit’s protection. Indeed, an unprotected single-qubit gate has been experimentally demonstrated in the soft 0-$\pi$ regime~\cite{gyenis2021}. Sufficiently fast and high-fidelity gates could help make these approaches more viable for larger-scale quantum computation.

This leaves an important gap. To our knowledge, no microwave-activated, unprotected two-qubit gate has been proposed for experimentally relevant soft \zps devices. Such gates are challenging because coupling two \zps qubits requires multi-node interactions, the computational states depend on external flux, and matrix elements are suppressed within the logical subspace due to protection. As a result, many gate mechanisms used in transmon architectures do not directly apply.

\begin{figure*}[!htbp]
\centering 
\includegraphics[width=1\textwidth]{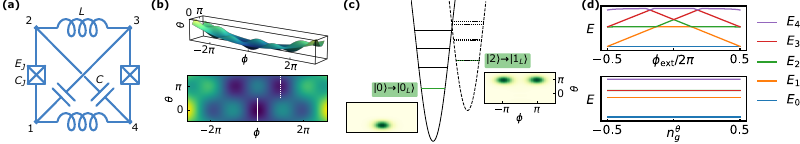}
\caption{An overview of a soft \zps qubit. (a) Circuit diagram. The circuit consists of two capacitors with capacitance $C$, two inductors with inductance $L$, and two Josephson junctions with Josephson energy $E_{J}$ and junction capacitance $C_J$. Each circuit node, labeled by $i\in\{1,2,3,4\}$, is associated with a flux variable $\varphi_i$ shown in \cref{eq:zp_transform}.
(b) Three-dimensional surface plot and two-dimensional map of potential energy $V(\theta, \phi)$ at $\phiext=0$, with parameters of qubit 1 from \cref{tab:parameter_zp}.  (c) Lowest ten eigenvalues (horizontal lines) are plotted in the double-well potential with a solid-line well ($\phi=0$) and a dashed-line well ($\phi=\pi$), corresponding to the solid and dashed white lines in (b). The two logical states $\ket{0_L} = \ket{0}$ and $\ket{1_L}=\ket{2}$ are highlighted in green and are strongly localized inside the two valleys.
(d) Energy spectrum of the lowest five eigenvalues versus external flux $\phiext/2\pi$ in flux quanta at $n_g^{\theta}=0$ in the top panel, and versus offset-charge $n_g^{\theta}$ at $\phiext=0$ in the bottom panel.
}\label{fig:circuit_potential}
\end{figure*}

In this article, we address this gap by proposing an unprotected controlled-$Z$ (CZ) gate for capacitively coupled soft \zps qubits. Since accurately simulating two coupled \zps circuits requires a large Hilbert space, we focus on the idealized zero-disorder limit. 
This assumption substantially reduces the required Hilbert-space dimension but eliminates an important source of device imperfections. The resulting gate performance should therefore be interpreted as a best-case bound for realistic devices.

The organization of the paper is as follows. In \cref{sec:basics}, we introduce the basics of a single \zps qubit. In \cref{sec:1q}, we optimize single-qubit $X$ gates using a voltage drive that capacitively couples to the $\phi$ or $\theta$ mode. 
In \cref{sec:coupling}, we introduce a capacitive coupling scheme between two \zps circuits. 
In \cref{sec:cz}, we introduce the CZ gate.
In \cref{sec:fidelity_2q}, we develop a two-qubit noise model and evaluate the performance of the CZ gate in the presence of noise.
Finally, in \cref{sec:conc}, we conclude with open questions.

\section{Basics of the \texorpdfstring{\zp}{zero pi} Qubit}\label{sec:basics}
In this section, we introduce the basics of the \zps qubit. Readers interested in more details should consult references~\cite{dempster2014,di2019,groszkowski2018,gyenis2021}.
The four-node \zps circuit is built using pairs of capacitors, inductors, and Josephson junctions. As shown in \cref{fig:circuit_potential}(a), each circuit node is connected to one element of each type.

 Because the \zps circuit contains multiple modes, the node flux variables $\varphi_i$ are not the easiest way to understand the circuit. To obtain variables that are easier to interpret, we use the variable transformation from Ref.~\cite{dempster2014}
\begin{align}\label{eq:zp_transform}
\left(\begin{array}{c}
{\theta} \\
{\phi} \\
{\zeta} \\
{\Sigma} \\
\end{array}\right)=\frac{1}{2}\left(\begin{array}{rrrr}
-1 & \phantom{-}1 & 1  & -1 \\
-1 & 1 & -1  & 1 \\
1 & 1 & -1  & -1 \\
1 & 1 & 1  & 1 \\
\end{array}\right)\left(\begin{array}{c}
\varphi_1 \\
\varphi_2 \\
\varphi_3 \\
\varphi_4 \\
\end{array}\right) \,.
\end{align}
The $\theta$ mode is transmon-like, as it depends on the phase differences across junctions and capacitors. Similarly, the $\phi$ mode is fluxonium-like, as it depends on the phase differences across junctions and inductors.

We make the simplifying assumption that all the capacitances, inductances, and junctions are the same (zero-disorder). Under this assumption, the LC-like $\zeta$ mode decouples exactly from other modes (and can be ignored), as can the cyclic $\Sigma$ mode~\cite{taylor2005}. The energy stored in any individual circuit element is thus parameterized by $E_C = e^2/2C$, $E_L = \phi_0^2/ L$, and $E_J = I_c \phi_0$, where $\phi_0 = \hbar /2e$ is the reduced flux quantum, $I_c$ is the critical current of the Josephson junction, and $C$ and $L$ are the capacitance and inductance, respectively. Because the zero-disorder assumption is not experimentally realistic, our work represents an upper bound on the performance for more realistic devices.

In the zero-disorder limit, we consider only the $\theta$ and $\phi$ modes. Circuit quantization yields generalized flux operators  $\hat{\theta}$ and $\hat{\phi}$ and conjugate charge operators $\hat{n}_{\theta}$ and $\hat{n}_{\phi}$ such that, e.g., $[\hat{\phi}, \hat{n}_\phi] = i$ \cite{dempster2014}. 
For the remainder of the paper, we will omit hats on operators for simplicity.
We also introduce an offset charge $n_g^{\theta}$ and an external flux bias $\phi_{\rm ext}$.
Using these operators, the two-mode quantum Hamiltonian for the \zps circuit is expressed as
\begin{align} \label{eq:zp Hamiltonian}
\begin{split}
    H_{0\text{-}\pi} =&\, 4E_C^\theta ({n}_{\theta}-n_g^{\theta})^2 + 4E_C^\phi {n}_{\phi}^2 \\    
    &+ E_L {\phi} ^2  - 2 E_J \cos{\theta} \cos\left({\phi} -\frac{\phiext}{2}\right) \,,
\end{split}
\end{align}
where $\phi_{\rm ext} = \Phi_{\rm ext}/\phi_0$. The charging energies of the $\theta$ and $\phi$ modes are
\begin{align}
    E_C^\theta = \frac{e^2}{2C_\theta} \quad \text{and} \quad 
    E_C^\phi  = \frac{e^2}{2C_\phi}  \, ,
\end{align}
where $C_\theta =2C +2C_J +C_0$ and $C_\phi = 2C_J +C_0$. Here $C_J$ is the junction capacitance and $C_0$ is the capacitance between the nodes and ground [not shown in \cref{fig:circuit_potential}(a)]. 

We now explain why the \zps is a protected qubit. Protection against energy relaxation is achieved by choosing logical states with disjoint support \cite{gyenis2021moving}. Any charge matrix elements (and corresponding decay rates) between them are thus suppressed. The selection of the computational states of the \zps circuit can thus be understood from the potential energy $V(\theta, \phi) = E_L \phi^2 - 2 E_J \cos\theta \cos\phi$, corresponding to the second line of \cref{eq:zp Hamiltonian}, evaluated at the working point $\phiext = 0$. 
In \cref{fig:circuit_potential}(b) we plot $V(\theta, \phi)$ using experimentally realized parameters of the \zps qubit. The exact values are shown as qubit 1 in \cref{tab:parameter_zp}. As seen in \cref{fig:circuit_potential}(b)--(c), the logical states of the soft \zp, $\ket{0_L}=\ket{0}$ and $\ket{1_L}=\ket{2}$, are chosen to be the lowest states in the $\phi=0$ and $\phi=\pi$ valleys. To ensure disjoint support, the effective barrier height between the two valleys ($\approx 4E_J$) must be much larger than the qubit frequency ($\approx E_L \pi^2$) and the kinetic energy ($E_C^{\theta}$) along the $\theta$ direction, i.e., $E_J\gg E_L, E_C^{\theta}$. Note that the first excited state is in the $\phi=0$ valley, so selecting $\ket{1_L} = \ket{1}$ would not yield a protected qubit.

\begin{table}[!htb]
\centering
\setlength{\tabcolsep}{3.5pt}
\begin{tabular}{cccccccc}
\hline \hline
Q & $E_J$ & $E_L$ & $E_C^{\theta}$ & $E_C^{\phi}$ & $E_{C_c}$ & $E_{C_0}$ & $g_{\theta}$ \rule{0pt}{2.8ex} \\[1pt] \hline
\rule{0pt}{2.4ex} 1   & 6.013  & \multirow{2}{*}{0.377}  & \multirow{2}{*}{0.092}  & \multirow{2}{*}{1.14} & \multirow{2}{*}{1.0} & \multirow{2}{*}{1.491} &  \multirow{2}{*}{0.031}    \\ 
2   & 5.412  &  &   &  &   & &    \\ \hline \hline
\end{tabular}
\caption{
Parameters for two soft \zps qubits in units of $h \cdot \text{GHz}$. Qubit 1 is the experimentally realized soft \zps from \cite{gyenis2021}, and qubit 2 is an otherwise identical system with $E_J$ reduced by 10\%. We consider qubit 1 for all single-qubit studies.
}\label{tab:parameter_zp}
\end{table}

The dephasing protection of the \zps is realized by suppressing the charge and flux dispersions. In \cref{fig:circuit_potential}(d) we plot the eigenvalues of \cref{eq:zp Hamiltonian} as a function of the offset charge $n_g^\theta$ and external flux $\phi_{\rm ext}$.
The charge dispersion is exponentially suppressed via a large ratio of $E_J/ E_C^{\theta}$, following the same mechanism as the transmon qubit \cite{koch2007}.
As for the flux dispersion, the symmetric and antisymmetric states in the two local minima of the $\theta=\pi$ valley give a hyperbolic dispersion as a function of $\phi_{\rm ext}$, so first-order protection can be achieved at $\phi_{\rm ext}=0$. 
Further flattening of the flux dispersion requires delocalization of wavefunctions along the $\phi$ direction, which entails larger kinetic energy $E_C^\phi$. To summarize, an ideal hard \zps qubit requires $E_L,E_C^\theta \ll E_J,E_C^\phi$ in order to achieve full protection against both energy relaxation noise and pure dephasing noise from flux and charge fluctuations.

\begin{figure}[!htbp]
\centering 
\includegraphics[width=1\columnwidth]{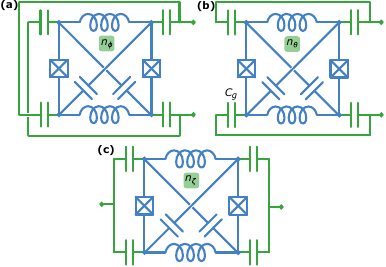}
\caption{Electrical connections required to couple to the $\phi$, $\theta$, and $\zeta$ modes of the \zps qubit~\cite{di2019}. The \zps and coupling circuit are shown in blue and green respectively. A microwave source is connected across the two green nodes to drive a specific mode.
}
\label{fig:circuit_3modes}
\end{figure}

To control the \zps qubit, we must couple to it. We consider only capacitive coupling, as inductive connections break the periodicity of the qubit potential -- except in the $L \to \infty$ limit~\cite{di2019}. We are primarily interested in coupling to the $\theta$ and $\phi$ modes, but coupling to the $\zeta$ mode may be necessary to cool the system and suppress photon shot noise \cite{di2019}.
To selectively couple to a single mode of the \zps qubit, individual connections to all four circuit nodes are required (see \cref{fig:circuit_3modes}), which is difficult to achieve in practice. Recent work has aimed to enable access to all four nodes~\cite{KimAPS2024,kim2025protected} through the use of parallel-plate instead of interdigitated capacitors. In this paper, we assume that each node of the \zps circuit can be electrically accessed and that the individual modes can be addressed.

In the experimental implementation of \zp, the authors use a higher non-logical state $\ket{9}$ to transform the qubit into an effective $\Lambda$ system and transfer population between the two logical states~\cite{gyenis2021}. Since certain nodes of the circuit were inaccessible, it was not possible to employ the coupling schemes in \cref{fig:circuit_3modes}. The experimental device induces transitions primarily by coupling to the $n_\phi$ operator, with residual coupling to $n_\theta$. 
This scheme is known to have problems, such as leakage into even higher energy levels~\cite{PremkumarAPS2023,premkumar2023hamiltonian}. We will discuss these limitations in detail in the next section. There are several alternative proposals for single-qubit gates for the \zps qubit, but they are primarily focused on the more strongly protected ``hard'' \zps regime \cite{di2019,kolesnikow2025}.

\section{Single-Qubit \texorpdfstring{$X$}{X} Gates}
\label{sec:1q}
In this section, we investigate the limitations of unprotected single-qubit $X$ gates in the soft \zps qubit that use population transfer via a non-logical state.
Assuming the capacitive coupling shown in \cref{fig:circuit_3modes}, we consider driving the $\phi$ and $\theta$ modes individually.
Because both gates are driven through the charge operators, we refer to them as the $n_\phi$-driven and $n_\theta$-driven gates.
We show that shorter, higher-fidelity operations are achieved by driving $n_\theta$, which is consistent with previous studies \cite{abdelhafez2020universal}.

\subsection{Driven Hamiltonian}\label{sec:1q_drive}
All single-qubit gate schemes in this work are implemented using two-tone drives.
The full Hamiltonian of the qubit with drive can be expressed as
\begin{align}\label{eq:H_total_and_drive_xgate}
    H_{\rm tot} = H_{0\text{-}\pi} + H_{\rm drive},
\end{align}
where the parameters for $H_{0\text{-}\pi}$ come from the first qubit in \cref{tab:parameter_zp}.
The microwave drives are parametrized as
\begin{align} \label{eq:zp_H_drive}
H_{\rm drive} = \sum_i {A_if(t)\cos{(\omega_{di} t)} n_{\rm op} } \, ,
\end{align} 
where $A_i$ is the drive amplitude, $\omega_{di}$ is the drive frequency, and  $n_{\rm op} \in \left\{n_{\rm \theta}, n_{\rm \phi}\right\}$ is a charge operator. For a pulse of length $t_g$, we choose $f(t)$ to be a Gaussian drive envelope\footnote{Because a Gaussian $g(t) = \exp[-(t-t_g/2)^2 / 2(t_g/4)^2] $ has $\mu=t_g/2$ and $\sigma = t_g/4$, the envelope $f(t) = \exp[-8t(t-t_g) / t_g^2] - 1 = e^2 (g(t) - g(0)) = e^2 (g(t) - g(t_g))$ has a $4\sigma$ width. Note that $f(t)$ has a maximum of $e^2-1$, so the maximum amplitude of the drive is $A_i f(t)= A_i(e^2-1)$.}
\begin{align} \label{eq:gauss_envelope}
f(t) = 
\begin{cases}
  0, &  t < 0\ \text{or}\  t > t_g \\
  e^{-8t(t-t_g)/t_g^2}-1 , &  0 \leq t \leq t_g
\end{cases} ~.
\end{align}

Below we identify an effective $\Lambda$ system for each charge operator by finding an intermediate state $\ket{\lambda}$ with achievable transitions $\ket{0_L} \leftrightarrow \ket{\lambda}$ and $\ket{1_L} \leftrightarrow \ket{\lambda}$. 
The transition frequency between levels $\ket{a}$ and $\ket{b}$ is denoted as $\omega_{a\text{-}b}$.
Each transition is driven by a separate pulse with amplitude $A_i$ and drive frequency 
\begin{equation}
  \omega_{di} =  \omega_{i\text{-} \lambda}+\delta_{i}\, ,
\end{equation} 
where $i \in \{ 0_L, 1_L\}$ and $\delta_{i}$ is the detuning.  Similarly, we denote the charge matrix element between levels $\ket{a}$ and $\ket{b}$ as 
\begin{equation}
    n_{\rm op}^{a\text{-}b} = \left|\bra{a} n_{\rm op} \ket{b}\right|.
\end{equation}

\subsection{\texorpdfstring{$X$}{X} gate by driving \texorpdfstring{$n_\phi$}{phi} } \label{sec:1q_phi} 

\begin{figure}[!htbp] 
    \includegraphics[width=1\columnwidth]{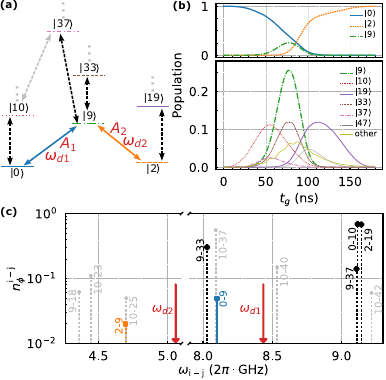}
\centering
\caption{
Numerically optimized $X$ gate by driving $n_\phi$, with $F= 99.659\%$ and $t_g=180.0~\mathrm{ns}$.
(a) Energy level diagram. 
The desired transitions, $\ket{0}$-$\ket{9}$ and $\ket{2}$-$\ket{9}$, are shown with blue and orange arrows. These transitions are driven with amplitudes $A_i$ and frequencies $\omega_{di}$ for $i\in\{1,2\}$.  Leakage pathways are denoted with dashed arrows. Black arrows represent ``level-1'' leakage transitions, while gray arrows represent ``level-2'' leakage transitions that involve level-1 states. 
(b) Population transfer. 
``other'' denotes the total population of states that are not explicitly shown. 
(c) Charge matrix elements $n_\phi^{i \text{-} j}$ versus transition frequency $\omega_{i \text{-} j}$.
The line colors and styles match those in (a).
The desired transitions have $\omega_{0\text{-}9}/2\pi=8.100\ {\rm GHz}$ and $\omega_{2\text{-}9}/2\pi=4.697\ {\rm GHz}$, with $n_{\phi}^{0\text{-}9}=0.049$ and $n_{\phi}^{2\text{-}9}=0.020$.
Simulations use the lowest 1000 levels; see \cref{app:1Qgate_simulation} and our open-source GitHub repository \cite{zp2q_repo} for numerical implementation details. Pulse parameters can be found in \cref{tab:xgate_parameters}. 
}\label{fig:xgate_phi}
\end{figure}

We first consider directly coupling to the $\phi$ mode. The $n_\phi$ operator induces an effective $\Lambda$ system involving $\ket{0}$, $\ket{2}$, and $\ket{9}$, as illustrated in \cref{fig:xgate_phi}(a). The colored arrows denote the target transitions. Numerous additional transitions give rise to significant leakage channels.
Black dashed arrows denote ``level-1'' leakage transitions that go from states in the $\Lambda$ system to level-1 states. Gray dashed arrows show level-2 leakage (from level-1 to level-2 states), and gray dotted lines indicate higher-level leakage. We use these connections to infer leakage pathways. For example, state $\ket{37}$ couples to states ($\ket{9},\ket{10}$), which in turn couple back to the computational space ($\ket{0},\ket{2}$).

\Cref{fig:xgate_phi}(b) shows the population of several levels during an $X$ gate of fidelity 99.659\% with gate time $t_g=180.0$ ns. The populations of logical states and leakage states are plotted separately, with $\ket{9}$ shown in both subplots for scale. We show only the most-populated leakage states, while the total population of all remaining states is grouped into the category “other.” Multiple leakage states have comparable population to state $\ket{9}$ during gate operation, with the top three being $\ket{10}$, $\ket{19}$, and $\ket{33}$.

To identify the associated leakage transitions, we plot charge matrix elements $n_\phi^{i \text{-} j}$ for these states as a function of transition frequency $\omega_{i \text{-} j}$ in \cref{fig:xgate_phi}(c).
The two transitions forming the $\Lambda$ system are shown in blue and orange, with corresponding drive frequencies marked by red arrows. These optimized drives are off-resonant due to the presence of nearby transitions.
The dashed lines indicate levels of leakage using the same color scheme as \cref{fig:xgate_phi}(a). The dominant level-1 leakage transitions appear to be $\ket{0}$-$\ket{10}$, $\ket{2}$-$\ket{19}$, and $\ket{9}$-$\ket{33}$.
Importantly, the $\ket{10} \leftrightarrow \ket{37}$ transition is nearly resonant with the $\ket{0} \leftrightarrow \ket{9}$ transition but has a much larger matrix element. Leakage to such a transition can be reduced by limiting population in the intermediate and level-1 leakage states. Due to frequency crowding, the drive strength must be kept small compared to the detuning from neighboring transitions to suppress leakage, resulting in long gate times.

\subsection{\texorpdfstring{$X$}{X} gate by driving \texorpdfstring{$n_\theta$}{theta}}  \label{sec:1q_theta}
\begin{figure}[!htbp] 
    \includegraphics[width=1\columnwidth]{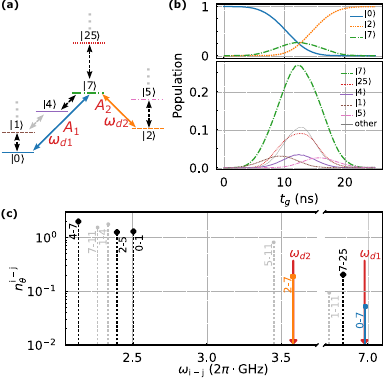}
\centering
\caption{
Numerically optimized $X$ gate by driving $n_\theta$, with $F=99.758\%$ and gate time $t_g$ = 25.0 ns.
(a) Energy level diagram. 
Leakage pathways are denoted with dashed arrows. Black arrows show ``level-1'' leakage transitions involving an intentionally populated state ($\ket{0}$, $\ket{2}$, or $\ket{7}$). 
(b) Population transfer. ``other'' includes the total population of all states not explicitly shown.
(c) Charge matrix elements $n_\theta^{i \text{-} j}$ versus transition frequency $\omega_{i \text{-} j}$.  
The two Raman transition frequencies are $\omega_{0\text{-}7}/2\pi=6.983\ {\rm GHz}$ and $\omega_{2\text{-}7}/2\pi=3.580\ {\rm GHz}$, with $n_{\theta}^{0\text{-}7}=0.052$ and $n_{\theta}^{2\text{-}7}=0.191$. Pulse parameters can be found in \cref{tab:xgate_parameters}.
}\label{fig:xgate_theta}
\end{figure}

Now we perform an $X$ gate by driving the $\theta$ mode. The $n_\theta$ operator induces a $\Lambda$-like system involving $\ket{0}$, $\ket{2}$, and $\ket{7}$. 
\Cref{fig:xgate_theta}(b) shows the population transfer for an $X$ gate with a fidelity of 99.758\% and gate time $t_g$ = 25.0 ns. The two drive frequencies $\omega_{d1}$ and $\omega_{d2}$ are indicated by red arrows in \cref{fig:xgate_theta}(c). Unlike the $n_\phi$ case, there is a single most prominent leakage state $\ket{25}$, which is populated through the $\ket{7}$-$\ket{25}$ transition. The next most prominent leakage states are $ \ket{1}$, $\ket{4}$, and $\ket{5}$. The transitions that populate these states are strongly detuned, but the relevant matrix elements are roughly an order of magnitude larger than those in the $\Lambda$ system, leading to significant population transfer. Nonetheless, there is substantially less leakage overall when driving the $\theta$ mode compared to driving the $\phi$ mode.
 
We can also compare our results with those of \citet{abdelhafez2020universal}.  In that work, the authors use optimal control to design an $X$ gate for the \zps qubit in a harder parameter regime (larger $E_J, E_C^\phi$ and smaller $E_L,E_C^\theta$).
The authors drive $n_\theta$ with an amplitude of 1.5 GHz $\approx 238\times 2\pi$ MHz, which is comparable to the drive amplitude used in our gate; see \cref{tab:xgate_parameters}.
For a gate time $t_g=$ 60 ns, they achieve an $X$ gate with 98.6\% coherent fidelity.\footnote{This is also consistent with \citet{di2019} when the fidelity and gate time are extracted from its results. That work uses the same parameters as Ref.~\cite{abdelhafez2020universal}.}
Although this indicates lower performance than for our $X$ gate, slower gates are expected in this more strongly protected \zps parameter regime. 
Indeed, when we simulate a Raman-style $X$ gate in the same parameter regime, we see comparable gate fidelities for this gate time.

\subsection{Single-Qubit Noise Model}\label{sec:noise_1q}
To study the effects of noise on these gates, we use a Lindblad master equation to phenomenologically include energy relaxation and dephasing. The evolution of the state matrix is given by
\begin{align} \label{eq:master_eq}
\frac{d\rho}{dt} = -\frac{i}{\hbar}[H, \rho] + \sum_{i,j} \mathcal{D}[L^r_{ij}]\rho + \sum_l \mathcal{D}[L^\varphi_l]\rho,
\end{align}
where $\mathcal{D}[X]\rho = X \rho X^\dagger - \frac{1}{2}( X^\dagger X \rho + \rho X^\dagger X)$ is the dissipation superoperator, and $L^r_{ij}$ and $L^\varphi_l$ are collapse operators for relaxation and dephasing, respectively.
The relaxation and dephasing operators have the form
\begin{subequations}\label{eq:lindblad_ops}
 \begin{align}
 L^r_{ij} = \sqrt{\frac{1}{T^r_{ij}}}  \ket{i} \bra{j}  ,\\
L^\varphi_l = \sqrt{\frac{2}{T^\varphi_l}}  \ket{l} \bra{l} \, ,
\end{align}
\end{subequations}
where $T_{ij}^r$ is the relaxation time of the transition $\ket{j} \rightarrow\ket {i}$ and $T^\varphi_l$ is the dephasing time of the state $\ket{l}$. 

The main challenge in building a noise model is specifying the operators $L^r_{ij}$ and $L^\varphi_l$. Our noise model is similar to the one considered in Ref.~\cite{abdelhafez2020universal}.
For energy relaxation, we consider the relevant charge operator $n_{\rm op}$ (with ${\rm op}\in\{\theta,\phi\}$ depending on the drive) and compute the downward transition rates between states $\ket{j} \rightarrow \ket{i}$ as
\begin{align} \label{eq:zp_t1}
\gamma_{j \text{-} i}^r = \Gamma \left| \bra{i} n_{\rm op} \ket{j} \right|^2.
\end{align}
The overall transition rate $\Gamma$ is not fixed by the model and must be set. Because both gate schemes use an effective $\Lambda$ system, we fix $\Gamma$ by assigning a target lifetime $T$ to the dominant decay channel from the intermediate state $\ket{\lambda}$.
Specifically, we identify the largest matrix element connecting $\ket{\lambda}$ to a lower level $\ket{M}$, and choose $\Gamma$ so that the corresponding decay time is $1/\gamma^r_{\lambda \text{-} M} = T$, where $T$ is a target value (e.g., $3,\ 30,\ 170~\mu\mathrm{s}$). Once $\Gamma$ is set, all other decay rates follow. For $n_\phi$, the intermediate state is $\ket{9}$, and the dominant transition is $\ket{9} \leftrightarrow \ket{4}$; we set $1/\gamma^r_{9 \text{-} 4} = T$. For $n_\theta$, the intermediate state is $\ket{7}$, and we fix $\Gamma$ using $1/\gamma^r_{7 \text{-} 4} = T$.

\begin{figure}[!htbp] 
\includegraphics[width=\columnwidth]{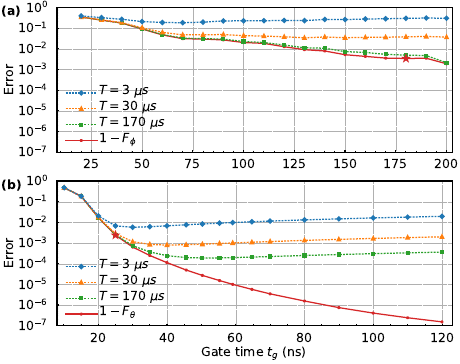}
\centering
\caption{
Simulated $X$ gate error $1-F$ versus gate time $t_g$ for numerically optimized pulses driving (a) $n_\phi$ and (b) $n_\theta$.
Solid lines represent coherent gate error without added noise. Dashed lines show the performance including open-system evolution with energy relaxation and pure dephasing, with $T_1=T_\phi=T\in\{3,30,170\}\,\mu\mathrm{s}$.
The two stars correspond to the pulses considered in \cref{fig:xgate_phi} and \cref{fig:xgate_theta}. 
The simulations here use a reduced model with 160 levels for the $n_\phi$ gate and 157 levels for the $n_\theta$ gate.
See \cref{app:1Qgate_simulation} for further numerical implementation details.
}\label{fig:xgate_fidelity}
\end{figure}

\begin{table*}[!htbp]
\centering
\setlength{\tabcolsep}{3.5pt}
\renewcommand{\arraystretch}{1.2} 
\begin{tabular}{ccccc ccccc cc }
\hline \hline
$n_{\rm op}$ & Intermediate & $ n_{ \rm op}^{ 0\text{-} i} $ & $n_{\rm op}^{ 2\text{-} i} $ & Leakage
& $t_g$ & $\Omega_{0\text{-}i}$  & $\Omega_{2\text{-}i}$  & $\delta_1$ & $\delta_2$ & $F_{\rm optimize}$ & $F_{\rm open}$   \\
& states &  &  & states & (ns) & ($2\pi\cdot$MHz)  & ($2\pi\cdot$MHz) & ($2\pi\cdot$MHz) & ($2\pi\cdot$MHz) &  & ($T$=30$\mu$s) \\
\hline 
$n_\phi$ & $\ket{9}$ & 0.049 & 0.020 & $[\ket{10}, \ket{19}, \ket{33} ]$  & 180.0 & 66.08 & 10.81 & 335.73 & 360.59 & 99.659\% & 96.183\%  \\ 
$n_\theta$ & $\ket{7}$ & 0.052 & 0.191 & $\ket{25}$ & 25.0  & 59.87 & 56.29  & -3.79  & 1.54   & 99.758\% & 99.713\% \\ 
\hline \hline
\end{tabular}
\caption{
Numerically optimized parameters corresponding to the two highlighted red stars in \cref{fig:xgate_fidelity}. The drive amplitudes and detunings $A_i/2\pi,\ \delta_i/2\pi$ are limited to be less than $500$ MHz. The effective Rabi rates are given by $\Omega_{0\text{-}i}=A_1(e^2-1)n_{\rm op}^{0\text{-}i}$ and $\Omega_{2\text{-}i}=A_2(e^2-1)n_{\rm op}^{2\text{-}i}$.
}
\label{tab:xgate_parameters}
\end{table*}

Charge dispersion is negligible at the chosen operating point [\cref{fig:circuit_potential}(d)], so we include only flux-induced pure dephasing. At $\phi_{\rm ext}=0$, the first-order flux sensitivity vanishes, and the dephasing rate is determined by the curvature of the transition frequency~\cite{hays2025nondegenerate},
\begin{align} \label{eq:zp_tphi}
\frac{1}{T^\varphi_{0\text{-}l}} = A_{\phi_{\rm ext}} \left| \frac{\partial^2 \omega_{0\text{-}l}}{\partial \phi_{\rm ext}^2} \right|.
\end{align}
We choose $A_{\phi_{\rm ext}}$ such that $T^\varphi_{0\text{-}2}=T$, after which \cref{eq:zp_tphi} determines the remaining dephasing times. We use the $\ket{0}\leftrightarrow\ket{2}$ transition for this calibration because it has the largest flux curvature among the transitions in the effective $\Lambda$ system.

There are two key points to emphasize. First, when we specify a noise strength such as $T = 3 \mu\text{s}$, we refer to both relaxation and dephasing times associated with the relevant transitions of the gate. Second, for simplicity, we adopt a family of noise models in which relaxation and dephasing rates are equal, i.e., $T^r = T^\varphi = T$.

\subsection{Noisy gate fidelity vs. gate time} \label{sec:fidelity_1q}
We evaluate gate performance using the average infidelity between the ideal $X$ gate and its noisy implementation $\tilde{X}$, denoted $1-F(X,\tilde{X})$~\cite{Pedersen2007,Wood2015}. 
For each gate duration $t_g$, we optimize the pulse ($A$, $\delta$) parameters to minimize $\log_{10}(1-F)$ in a closed quantum system. For additional details, see \cref{app:sec:model_reduce_1q,app:sub_sec_pluse_opt}.

The optimization pipeline is as follows. 
For each gate time $t_g$, we optimize the two amplitudes $A_{1,2}$ and detunings $\delta_{1,2}$ in the closed-system model. The amplitudes are constrained to $A/2\pi\in[0,500)~\text{MHz}$, and the detunings $\delta=\omega_d-\omega_{i\text{-}j}$ are varied within $\pm\,2\pi\times500~\text{MHz}$ around the target transition $\omega_{i\text{-}j}$.
The gate error is then computed by applying pulses optimized in the closed-system model to the noisy evolution generated by the master equation in \cref{eq:master_eq}. For more details, see \cref{app:sub_sec_pluse_opt,app:sub_sec_noisy_1Q}.

\Cref{fig:xgate_fidelity}(a) shows the simulated $X$-gate error $1-F$ versus gate time $t_g$ when driving $n_\phi$, while \cref{fig:xgate_fidelity}(b) shows the corresponding results for driving $n_\theta$. The $y$-axis scales are matched to enable direct comparison. The key observation is that $X$ gates implemented via $n_\theta$ driving achieve both shorter gate times and higher fidelities than those obtained with $n_\phi$ driving. 

There are two main reasons for this advantage. First, in our parameter regime the relevant matrix elements of $n_\theta$ are significantly larger than those of $n_\phi$ (see \cref{tab:xgate_parameters}). In particular, $n_\theta^{2\text{-}7}/n_\phi^{2\text{-}9}\sim 10$ is an order of magnitude larger than $n_\theta^{0\text{-}7}/n_\phi^{0\text{-}9}\sim 1$. 
Second, the $n_\theta$ drive has fewer leakage pathways than the $n_\phi$ drive, as seen in \cref{fig:xgate_phi}(a)–(b) and \cref{fig:xgate_theta}(a)–(b). 
With fewer leakage channels, stronger and shorter drives can be used without compromising fidelity. Moreover, the optimized drive frequencies for $n_\phi$ are often off-resonant (for the gate times considered), while those for $n_\theta$ are typically near-resonant, further reducing leakage.

The optimized gates are significantly faster than the experimentally demonstrated gate in Ref.~\cite{gyenis2021}.
In \cref{sec:x_gate_reproduce}, we give some explanation for this apparent discrepancy (e.g., experimental limitations on drive amplitude, mixed mode coupling, and decoherence not predicted by our phenomenological noise model). Ultimately, we show that our numerical models can roughly reproduce the behavior observed in the experiment.

\section{ Coupling for Two \texorpdfstring{\zp}{zero pi} Qubits} \label{sec:coupling}
Our single-qubit analysis shows that coupling to the $\theta$ mode provides larger charge matrix elements and fewer leakage pathways than coupling to the $\phi$ mode. Motivated by this finding, we focus on designing two-qubit gates based on a charge-charge interaction between the two circuits' $\theta$ modes, i.e., an $n_{\theta1} n_{\theta2}$ interaction.

\begin{figure}[!htbp] 
    \includegraphics[width=1\columnwidth]{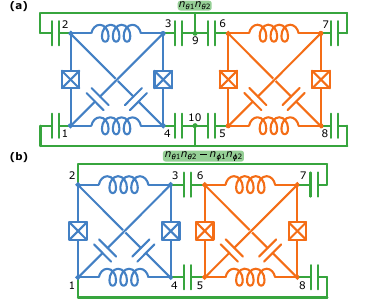}
\centering
\caption{
Two coupled \zps qubits with (a) pure $g_{\theta} n_{\theta1} n_{\theta2}$ and (b) mixed $g_{\theta} n_{\theta1} n_{\theta2} - g_{\phi} n_{\phi1} n_{\phi2}$ interaction terms. The individual qubits are shown in blue and orange, while the coupling circuit is drawn in green.
}\label{fig:circuit_2qubits}
\end{figure}

The $\theta$ mode involves all nodes of the \zps circuit, so producing an $n_{\theta1} n_{\theta2}$ interaction requires multi-node coupling. We show two such schemes in \cref{fig:circuit_2qubits}. The Hamiltonian for these two \zps systems can be written as
\begin{align}\label{eq:coupled_zp_ham}
     H_{\rm cp} =  H_{0\text{-}\pi}^{(1)} +  H_{0\text{-}\pi}^{(2)} +  H_{\rm int},
\end{align}
where the superscript $(i)$ refers to the $i$th qubit and $H_{\rm int}$ is the interaction Hamiltonian. The two circuits in \cref{fig:circuit_2qubits} (a) and (b) yield interactions
\begin{subequations}
\begin{align}
    H_{\rm int}^{\rm (a)} &= g_{\theta} n_{\theta1} n_{\theta2} ~, \\
    H_{\rm int}^{\rm (b)} & = g_{\theta} n_{\theta1} n_{\theta2} - g_{\phi} n_{\phi1} n_{\phi2} ~.
\end{align}
\end{subequations}
Circuit (a) implements a pure $n_{\theta1} n_{\theta2}$ interaction and forms the foundation of our two-qubit gate implementation. 
Circuit (b) does not produce a pure  $n_{\theta1} n_{\theta2}$ interaction.  Nonetheless, circuit (b) does have some advantages despite its mixed interaction. Notably, it requires fewer capacitors, making it easier to construct, and the renormalized $\tilde E_C^\phi$ remains light even with a large coupling capacitor.
A full derivation of both interactions is provided in \cref{app:quantize}. Since the remainder of the manuscript focuses on circuit~(a), we briefly outline its derivation below.

Because the coupling network in circuit~(a) is purely capacitive, it modifies only the kinetic terms in the Lagrangian. As labeled in \cref{fig:circuit_2qubits} (a), nodes 1–4 correspond to qubit 1, nodes 5–8 to qubit 2, and nodes 9–10 to the coupler.  
With this convention, the interaction Lagrangian in node flux variables is
\begin{align} \label{eq:zp_interact_theta_lagrangian_1}
T_{\rm{int}}=\frac{C_{c}}{2} \big [
&(\dot{\varphi }_1-\dot{\varphi }_{10}){}^2 
+(\dot{\varphi }_4-\dot{\varphi }_{10}){}^2      
+(\dot{\varphi }_5-\dot{\varphi }_{10}){}^2     \nonumber\\
&+(\dot{\varphi }_8-\dot{\varphi }_{10}){}^2      
+(\dot{\varphi }_2-\dot{\varphi }_9){}^2 
+(\dot{\varphi }_3-\dot{\varphi }_9){}^2        \nonumber\\
&+(\dot{\varphi }_6-\dot{\varphi }_9){}^2 
+(\dot{\varphi }_7-\dot{\varphi }_9){}^2 \big ]   \,.   
\end{align}
Here, $C_c$ is the value of the coupling capacitors, shown in green in \cref{fig:circuit_2qubits}. \Cref{eq:zp_interact_theta_lagrangian_1} shows that both qubits couple to nodes 9 and 10.

After performing the variable transformation in \cref{eq:zp_transform} and setting $\varphi_{\pm}=\varphi_9 \pm \varphi_{10}$, we obtain
\begin{align} \label{eq:zp_interact_theta_lagrangian_2}
T_{\rm{int}}
=&\frac{C_c}{2}\sum_{i=1}^{2} \big [\dot{\phi}_i^2 + \dot{\zeta}_i^2  
+(\dot{\varphi }_+ - \dot{\Sigma }_i)^2 
+(\dot{\varphi }_- - \dot{\theta }_i)^2 \big ] \, .
\end{align}
The $\varphi_-$ mode is coupled directly to both $\theta_1$ and $\theta_2$. There are no potential energy terms corresponding to the $\Sigma_1$, $\Sigma_2$, $\varphi_+$, and $\varphi_-$ modes, so they are ``free modes'' and can be removed from the Lagrangian \cite{Groszkowski2021scqubitspython,DingKuShiZhao2021,chitta2022computer}.

After free mode removal and circuit quantization, the coupling Hamiltonian is 
\begin{align} \label{eq:zp_2_qubit_coupling_theta_2}
 H_{\rm{int}} = 
g_\theta\, {n}_{\theta 1} {n}_{\theta 2} \, ,
\end{align}
with coupling energy 
\begin{equation} \label{eq:g_theta}
g_\theta = 
\frac{2e^2C_c}{\tilde C_{\theta} (\tilde C_{\theta} - C_c)} \approx \frac{e^2C_c}{2C^2}\,.
\end{equation} 
Here $\tilde C_{\theta}=2C + 2C_J +C_0 +C_c = C_{\theta} + C_c $. The approximation is valid when $C\gg C_c, C_J, C_0$.  

One consequence of capacitive coupling is the renormalization of each qubit’s charging energy. For the circuit in \cref{fig:circuit_2qubits} (a), the charging energies become
 \begin{align}
    \tilde E_C^\phi = \frac{e^2}{2 \tilde C_\phi} , \quad 
    \tilde E_C^\theta = \frac{e^2(2 \tilde C_\theta-C_c)}{4 
\tilde C_{\theta}( \tilde C_{\theta} - C_c)}
     \, ,
 \end{align}
where $\tilde C_\phi = 2 C_J +C_0 +C_c = C_\phi +C_c$. The coupling capacitance makes both modes heavier ($\tilde E_C^\phi < E_C^\phi$ and $\tilde E_C^\theta  < E_C^\theta$), and these changes can be significant. Using the parameters from \cref{tab:parameter_zp}, $(C_J,C_0,C_c)\approx(2,13,19)\,\mathrm{fF}$, $\tilde E_C^\phi$ and $\tilde E_C^\theta$ decrease by approximately $50\%$ and $4\%$, respectively. The hard \zps regime requires $E_L,E_C^\theta\ll E_C^\phi$. This is difficult to achieve in circuit (a) because the coupling capacitance suppresses $E_C^\phi$.
Specifically, in a hard \zps regime, we expect $C_c\gg C_J$ and $C_0 \rightarrow 0$, so the $\phi$ mode charging energy scales with $\tilde E_C^\phi\propto 1/2C_c$. Thus, a very large inductance and shunt capacitance would be required to achieve the hard \zps regime.
Circuit (b) avoids this limitation because $\tilde E_C^\phi \propto \left[1/C_J+1/(C_J+C_c)\right]$, which approaches $1/C_J$ when $C_c\gg C_J$.

The coupling schemes in \cref{fig:circuit_2qubits} are experimentally demanding. They require capacitive access to all four nodes of each \zps qubit, together with a multi-node coupling network and additional control and readout circuitry. The resulting layout is substantially more complex than that of conventional superconducting-qubit architectures. Suppressing parasitic capacitances remains an open challenge. Our analysis also assumes identical circuit elements, for which unwanted coupling terms cancel exactly. In a realistic device, disorder in the coupling capacitors and stray capacitances will break this symmetry and generate residual interactions. We nevertheless adopt the zero-disorder limit because it reduces the Hilbert-space dimension, simplifies the numerical analysis, and establishes a best-case benchmark for gate performance. A detailed treatment of disorder, parasitic couplings, and layout constraints is left for future work.

Before closing this section, we briefly comment on generalizations of this coupling scheme.  Similar layouts can be used to connect any pair of \zps modes, and we show the required connections in \cref{tab:coupling_table}. As with the system in \cref{fig:circuit_2qubits} (a), these coupling circuits require many more capacitors than are typically needed in transmon architectures, even before accounting for drive lines.

\begin{table}[!htb]
\centering
\setlength{\tabcolsep}{7pt}
\renewcommand{\arraystretch}{1.05}
\begin{tabular}{ccc}
\hline \hline
\small
Interaction                    & Nodes coupled to 9        & Nodes coupled to 10     \\ \hline
$n_{\theta 1}n_{\theta 2}$  & 2,3 + 6,7         & 1,4 + 5,8       \\ 
$n_{\phi 1}n_{\phi 2}$      & 1,3 + 5,7         & 2,4 + 6,8       \\ 
$n_{\theta 1}n_{\phi 2}$    & 2,3 + 5,7         & 1,4 + 6,8       \\ 
$n_{\phi 1}n_{\theta 2}$    & 1,3 + 6,7         & 2,4 + 5,8       \\ \hline \hline
\end{tabular}
\caption{
Capacitive connections that couple the specified modes of two \zps qubits. Node numbers match \cref{fig:circuit_2qubits} (a). 
}\label{tab:coupling_table}
\end{table}

\section{Microwave-activated CZ gate}\label{sec:cz}
We now develop a CZ gate for two coupled \zps qubits, using the coupling scheme from \cref{sec:coupling} and a microwave drive applied to one qubit. 
Inspired by similar CZ and CCZ gates in fluxonium~\cite{nesterov2018microwave,ficheux2021fast, mazhorin2026nativecczgatefluxonium}, the gate drives a transition between a computational state and a higher non-computational state to accumulate a $\pi$ phase.
As in fluxonium, the \zps spectrum is crowded, necessitating careful consideration of competing transitions when optimizing gate performance.

\subsection{Driven Hamiltonian}\label{sec_sub:cz_drive_params}
The Hamiltonian of the coupled qubits with drive is
\begin{align}\label{eq:H_total_and_drive_cz}
    H_{\rm tot} = H_{0\text{-}\pi}^{(1)} + H_{0\text{-}\pi}^{(2)} +H_{\rm int} +H_{\rm drive}^{(2)},
\end{align}
where $H_{\rm drive}^{(2)}$ is a microwave drive on $n_{\theta 2}$ of the form given in \cref{eq:zp_H_drive}. The full circuit diagram is shown in  \cref{fig:cz_ntheta}(a), with the drive circuitry highlighted in red.
It should be noted that the capacitive coupling to the voltage source induces additional energy renormalization, which we do not consider, as it is well studied in prior work~\cite{di2019}.

We set device parameters so that both \zps circuits are in the soft regime, while allowing slight asymmetry.  
The first qubit, described by $H_{0\text{-}\pi}^{(1)}$, uses the same parameters as the experimental device in Ref.~\cite{gyenis2021}. 
The second qubit, $H_{0\text{-}\pi}^{(2)}$, has $E_J$ reduced by 10\%, with all other parameters unchanged (see \cref{tab:parameter_zp}).  

To analyze the gate operation, we establish the following notation.  The eigenstates of the uncoupled system ($H_{\rm int} = 0$) are denoted $\ket{k,l}_0$, where $k$ and $l$ label the eigenstates of the first and second qubits, respectively. These are commonly referred to as bare states. 
With finite coupling ($H_{\rm int}\ne 0$), the ``dressed'' eigenstates are denoted $\ket{k,l}$ with eigenvalues $E_{k,l}$. Each dressed state is indexed by the bare state with which it has the largest overlap. For example, the dressed state $\ket{5,0}$ is defined by its large overlap with $\ket{5,0}_0$: $\left|{}_0\!\langle 5,0| 5,0 \rangle \right|=0.91$.
In addition, a key quantity is the transition frequency between dressed states.  
For a transition $\ket{k,l}\text{-}\ket{k',l'}$, we denote the transition frequency as  $\omega_{k,l\text{-} k',l'} = |\omega_{k,l}-\omega_{k',l'}|$ with $\omega_{k,l}= E_{k,l}/\hbar$.

\subsection{Gate Concept and Transition Landscape}\label{sec:cz_gate_transition}
A two-qubit gate requires an interaction between the two \zps circuits, as described by the total Hamiltonian in \cref{eq:H_total_and_drive_cz}. Because charge matrix elements within the logical subspace are strongly suppressed, direct microwave-driven transitions between logical states are impractical. To overcome this limitation, we use $H_{\rm drive}^{(2)}$ to drive a transition between a logical state and a non-computational state, thereby avoiding direct population transfer within the logical subspace.

Many CPHASE-gate schemes selectively couple the computational state $\ket{1,1}_L$ to a non-computational auxiliary state, e.g.,  $\ket{2,0}$ for a transmon. The drive returns the population to $\ket{1,1}_L$ along a closed trajectory on the corresponding Bloch sphere. This cyclic evolution gives $\ket{1,1}_L$ a geometric phase relative to the other computational states, set by the solid angle enclosed by the trajectory~\cite{nesterov2018microwave,ficheux2021fast}. A complete resonant round trip produces a phase of $\pi$ and therefore implements a CZ gate. Closed trajectories that enclose a smaller solid angle produce a smaller controlled phase.

Our two-qubit gates exploit the static qubit--qubit interaction $H_{\rm int}$ to lift the degeneracy of otherwise identical transition frequencies, while $H_{\rm drive}^{(2)}$ selectively drives one of the resulting transitions.
For two coupled quantum systems, as the interaction strength $g$ is increased from 0, degenerate states hybridize if they are coupled by the interaction. The hybridization gap grows with the magnitude of the matrix element $\left|{}_0\bra{n',m'} H_{\rm int} \ket{n,m}_0\right|$. Corresponding transitions in the dressed basis are split. This splitting enables gates that selectively drive one of the previously degenerate transitions.\footnote{There is no direct coupling between logical states in the \zp, so the static $ZZ$ rate $(\omega_{0,2-2,2} - \omega_{0,0-2,0})/2\pi = \Gamma_{\rm ZZ}=171$ Hz of the coupled \zps system is substantially smaller than in transmon or fluxonium devices with comparable interaction strength.}

\begin{figure}[!htbp] 
    \includegraphics[width=1\columnwidth]{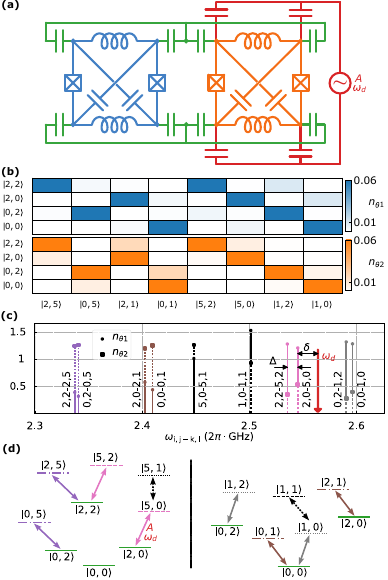}
\centering
\caption {  
CZ gate implemented by driving $n_{\theta 2}$.
(a) Circuit diagram. The coupling and drive circuits are depicted in green and red, respectively.  
(b) Dressed basis charge matrix elements $\bra{i,j}n_{\theta r}\ket{k,l}$ for $r\in \{1, 2\}$. The matrix elements are between two-qubit logical states (y-axis) and some non-logical states (x-axis). Four pairs of dominant matrix elements are present in both diagrams. 
(c) Charge matrix elements $\bra{i,j}n_{\theta r}\ket{k,l}$ for $r\in \{1, 2\}$ versus transition frequency $\omega_{i,j\text{-}k,l}$. 
Solid lines ending in circles denote $n_{\theta 1}$, and dashed lines ending in squares denote $n_{\theta 2}$.
(d) Energy level diagram. Four transition pairs in the CZ gate are denoted by solid arrows, using the same color scheme as in (c). Other leakage transitions are represented by dashed arrows. 
}\label{fig:cz_ntheta}
\end{figure}

We now identify a suitable transition for implementing the CZ gate. 
In the case of \zp, the interaction Hamiltonian most strongly couples states involving the logical states and the $\ket{1}$ and $\ket{5}$ non-computational states. As a result, four prominent pairs of conditionally split transitions emerge, which we show in \cref{fig:cz_ntheta} (c) and (d):
\begin{itemize}
    \item pink pair: $\ket{2,0}\leftrightarrow\ket{5,0}$ and $\ket{2,2}\leftrightarrow\ket{5,2}$,
    \item  gray pair: $\ket{0,0}\leftrightarrow\ket{1,0}$ and $\ket{0,2}\leftrightarrow\ket{1,2}$,
        \item purple pair: $\ket{0,2}\leftrightarrow\ket{0,5}$ and $\ket{2,2}\leftrightarrow\ket{2,5}$,
    \item brown pair: $\ket{0,0}\leftrightarrow\ket{0,1}$ and $\ket{2,0}\leftrightarrow\ket{2,1}$.
\end{itemize}
Recall that the physical state $\ket{2}$ is our logical $\ket{1}_L$. Only the transition $\ket{2,2}\leftrightarrow\ket{2,5}$ is analogous to the $\ket{1,1}_L \leftrightarrow \ket{2,0}$ transition used in a transmon CZ. 
However, all such transitions result in a gate equivalent to a controlled-phase gate up to local single-qubit rotations.

To find the most suitable candidate, we additionally consider the frequency mismatch $\Delta$, defined as the frequency difference between the two transitions within each pair. We expect that the most spectrally isolated transition will be the easiest to selectively drive. Among the four pairs, the pair $[\ket{2,2} \text{-} \ket{5,2}, \ket{2,0} \text{-} \ket{5,0}]$ has the largest $\Delta = | \omega_{2,0 \text{-} 5,0} - \omega_{2,2 \text{-} 5,2} |$.

The matrix elements of these transitions depend strongly on which qubit is driven. Transitions that primarily change the state of the first qubit, such as $\ket{2,0}\leftrightarrow\ket{5,0}$, have smaller matrix elements under an $n_{\theta2}$ drive than under an $n_{\theta1}$ drive.
This behavior is illustrated in \cref{fig:cz_ntheta}(c).
For the selected pair, the $\ket{2,0}\leftrightarrow\ket{5,0}$ transition is sufficiently strong, whereas its conditional counterpart $\ket{2,2}\leftrightarrow\ket{5,2}$ is weaker. The resulting matrix element asymmetry is greater than for an $n_{\theta1}$ drive, making this transition pair well suited for selective excitation.
We therefore choose $\ket{2,0}\leftrightarrow\ket{5,0}$ as the primary driven transition under an $n_{\theta2}$ drive.  The combination of a large conditional frequency splitting, an appreciable target matrix element, and a strong matrix element asymmetry makes this transition the most favorable choice for implementing the CZ gate.

In practice, we use the selected transition frequency as a starting point and numerically optimize pulse parameters to maximize gate fidelity. As seen in \cref{fig:cz_population}, the optimized drive activates multiple transitions, rather than simply driving a round trip between $\ket{2,0}$ and $\ket {5,0}$. It is possible to recover this simpler scheme at the cost of increased gate time. Regardless of the intermediate populations, the final ``candidate'' gate is approximately diagonal in the computational basis and has the form
\begin{equation}\label{eq:CZ_with_local_phases}
U_{\rm can} =  \mathrm{diag}\left[ e^{i \phi_{0,0}}, e^{i \phi_{0,2}}, e^{i \phi_{2,0}}, e^{i \phi_{2,2}}\right]\,.
\end{equation}
To arrive at the standard $ \text{CZ} =\mathrm{diag}[1,1,1,-1]$ gate, we apply single-qubit $z$ rotations  $R_z(\phi)=\exp[-i \phi Z/2] $ to \cref{eq:CZ_with_local_phases}. In particular, we transform $U_{\rm can} $ to  $U_{\rm std} = \left[ R_z(\alpha_2) \otimes R_z(\alpha_1) \right] U_{\rm can} $, where $\alpha_1 = \phi_{0,0}-\phi_{0,2}$ and $\alpha_2 = \phi_{0,0}-\phi_{2,0}$ cancel the local phases.
When applying these phase corrections to $U_{\rm can}$, we arrive at 
\begin{equation}
U_{\rm std} = e^{i \phi_g}\mathrm{diag} \left[ 1, 1, 1, e^{i\chi} \right] \, .
\end{equation}
Here, $\phi_g= (\phi_{0,2} +\phi_{2,0})/2$ is a global phase, and $\chi = \phi_{2,2}+ \phi_{0,0} - \phi_{0,2}- \phi_{2,0}$ is the controlled phase. For an ideal CZ gate, $\chi=\pi$.

We now briefly discuss an alternative drive scheme that results in a CNOT gate. This CNOT gate uses a two-qubit $\Lambda$-type transition, inspired by the single-qubit $X$ gates in \cref{sec:1q}. Thus it requires a two tone drive on one of the qubits. Full details are provided in \cref{sec:cnot}.
In noiseless simulations, the CNOT gate achieves fidelities above $99.9\%$, but it is about two times slower than the CZ gate at comparable fidelities. The slower operation results from the weaker transition matrix elements and smaller conditional frequency splittings involved. This CNOT construction serves as a proof of principle that the coupling architecture supports microwave-activated two-qubit gates beyond the CZ gate.

\section{CZ Gate Performance with Decoherence} \label{sec:fidelity_2q}
In this section, we present numerically optimized CZ pulses. We additionally extend the single-qubit noise model presented in \cref{sec:noise_1q} to two coupled qubits and examine the optimized pulses subject to this noise model. Performing these open-system simulations is computationally intensive and requires careful construction of the two-qubit Hamiltonian. We refer the interested reader to \cref{sec:2Qgate_simulation} for additional numerical details.

\begin{table*}[!htbp]
\centering
\setlength{\tabcolsep}{3.5pt}
\renewcommand{\arraystretch}{1.2} 
\begin{tabular}{cccc ccc cc}
\hline \hline
Gate & $n_{\rm op}$ & Transitions & Associated $n_{\rm op}$ & $t_g$ & $\Omega_1$   & $\delta_1$  & $F_{\rm optimize}$ & $F_{\rm open}$  \\ 
 &  &  &  & (ns) & ($2\pi\cdot$MHz)  & ($2\pi\cdot$MHz)  &  & ($T=30~\mu\mathrm{s}$) \\ 
\hline 
\multirow{2}{*}{CZ} & \multirow{2}{*}{$n_{\theta 2}$} & \multirow{2}{*}{$\ket{2,0}$-$\ket{5,0}$} & \multirow{2}{*}{$0.537$} & 91.95 & 61.45  & 25.19  & 99.893\% & 99.830\%  \\ 
 &  &  &  & 182.01 & 27.04  & 13.13  & 99.993\% & 99.908\%  \\ 
\hline \hline
\end{tabular}
\caption{
The parameters of the two CZ pulses corresponding to the red star and pentagon in \cref{fig:2q_fidelity}. $A_i/2\pi$
is limited to be less than $250$ MHz. 
The effective Rabi rate is given by $\Omega_{1}=A_1(e^2-1)n_{\rm op}$.
}\label{tab:parameter_cz}
\end{table*}

\subsection{Two-Qubit Noise Model}\label{sec:noise_2q}

Our phenomenological two-qubit noise model is built as a tensor product of independent single-qubit noise channels. Let $L^r_{ij}$ and $L^\varphi_l$ denote single-qubit relaxation and dephasing Lindblad operators from \cref{eq:lindblad_ops}.
The two-qubit operators are constructed as 
\begin{equation}
\mathfrak{L}^{(1)} = L \otimes I\,\, {\rm and} \,\, \mathfrak{L}^{(2)} = I \otimes L\,, 
\end{equation}
with $L \in \{L^r_{ij},\, L^\varphi_l\}$. Though each operator requires multiple indices to characterize the noise process, here we omit them for clarity. The decay and dephasing rates are chosen as in the single-qubit case (see \cref{sec:noise_1q}). In particular, we choose $T_1= T_\phi=T$ for both qubits. Dynamics are simulated using the master equation in \cref{eq:master_eq}. The full two-qubit state is evolved to obtain a propagator and gate fidelity. We provide a more thorough description of the open-system model in \cref{app:subsec_2Q_noise_details_and_converge}.

\subsection{Optimized Gate}\label{sec:2q_gate}

\begin{figure}[!htbp] 
    \includegraphics[width=1\columnwidth]{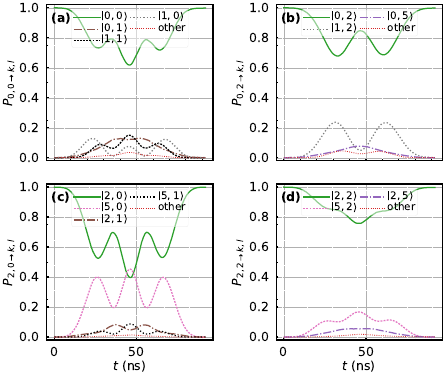}
\centering
\caption 
{Noiseless simulation of population transfer during a CZ gate at $t_{\rm g}=91.95$ ns with $F=99.893\%$.
$P_{i,j \text{-} k,l }$ denotes the population of state $\ket{k,l}$ for an initial state $\ket{i,j}$. The initial states in panels (a)--(d) are $\ket{0,0}$, $\ket{0,2}$, $\ket{2,0}$, and  $\ket{2,2}$, respectively.
The ``other'' state refers to the total population of all states not explicitly shown within the lowest 1000 dressed states.
}\label{fig:cz_population}
\end{figure}

To understand the gate mechanism, we examine the noiseless population dynamics during an optimized CZ gate. In \cref{fig:cz_population}, we show these dynamics for a gate with $t_g = 91.95$~ns and $F= 99.893\%$. The gate uses a single Gaussian pulse to simultaneously drive multiple transitions. The main CZ transition $\ket{2,0}\text{-}\ket{5,0}$, shown in panel (c), reaches approximately   $40\%$ population in $\ket{5,0}$. Additional non-computational states associated with the four identified transition pairs are also appreciably populated. For example, \cref{fig:cz_population}(d) shows population transfer from $\ket{2,2}$ to both $\ket{2,5}$ and $\ket{5,2}$.
Despite the nontrivial intermediate dynamics, leakage is small at the end of the gate, with the final population in all other states below $0.4\%$. These excursions are acceptable under the noise model considered in \cref{sec:noise_2q}, although intermediate populations can be reduced by increasing the gate duration or utilizing smaller controlled-phase operations.

\begin{figure}[!thbp] 
\includegraphics[width=0.99\columnwidth]{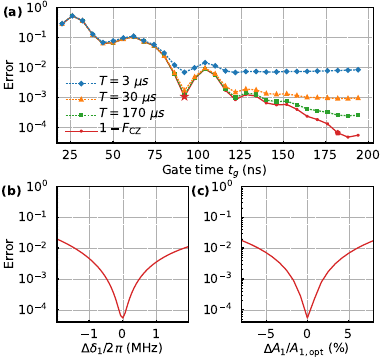}
\centering
\caption{ 
CZ gate performance. (a) Gate error ($1-F$) versus gate time $t_g$, using optimized pulse parameters for each $t_g$. Solid lines show errors for unitary (closed-system) evolution, while dashed lines include decoherence with $T_1=T_\phi=T\in\{3,30,170\}\,\mu\mathrm{s}$. The star marks the operating point illustrated in \cref{fig:cz_population}. We use a reduced model with 60 states to simulate this gate. See \cref{app:subsec_2Q_noise_details_and_converge} for additional modeling details. 
(b) Coherent gate error as a function of the drive-detuning deviation $\Delta\delta_1=\delta_1-\delta_{1,\mathrm{opt}}$ at $t_g=182$ ns, with the drive amplitude fixed at its optimized value. 
(c) Coherent gate error as a function of the relative drive-amplitude deviation $\Delta A_1/A_{\mathrm{opt}}=(A_1-A_{1,\mathrm{opt}})/A_{1,\mathrm{opt}}$ at $t_g=182$ ns, with the detuning fixed at its optimized value.
Full pulse parameters for both highlighted pulses are given in \cref{tab:parameter_cz}. 
}\label{fig:2q_fidelity}
\end{figure}

\Cref{fig:2q_fidelity} shows how the gate error $1-F$ scales with gate time $t_g$ for the CZ gate under different coherence times.
Panel (a) clearly shows that the CZ gate is coherence limited for shorter lifetimes. For $T=3\,\mu\mathrm{s}$, for example, the minimum error is approximately $10^{-2}$, whereas for $T=30\,\mu\mathrm{s}$ it reaches approximately $10^{-3}$. Within our phenomenological noise model, the dominant error arises from dephasing of the logical state $\ket{1}_L=\ket{2}$.
Panels (b) and (c) show the coherent gate-error landscapes near the optimized operating point as the drive detuning and amplitude are varied. This allows us to assess the robustness of gates to control-parameter errors. The coherent fidelity remains above $99.9\%$ for drive-frequency deviations of approximately $500$ kHz and drive-amplitude deviations of about $\pm2\%$.
The resulting coherent errors are much smaller than those introduced by dissipation, indicating that the gate is reasonably robust to calibration errors. For experimentally relevant coherence times, fidelity is therefore limited by logical-qubit coherence, rather than by coherent control errors.

Finally, we investigate the robustness of the optimized CZ gate to device-fabrication imperfections. We independently vary each circuit parameter of the second qubit by up to $\pm5\%$, re-optimize the gate, and compute the resulting infidelity. The results, summarized in \cref{app:sensitivity}, show that the gate is most sensitive to variations in $E_{C}$, $E_{L}$, and $E_{J}$, but relatively insensitive to changes in $\phi_{\mathrm{ext}}$, $n_{g}^\theta$, $E_{Cc}$, $E_{CJ}$, and $E_{C0}$. Nevertheless, the median degradation in gate performance is small over the parameter ranges considered, particularly for longer gate times.

\section{Conclusion}\label{sec:conc}

In summary, we have proposed and analyzed a two-qubit CZ gate for capacitively coupled soft \zps circuits.
Using a phenomenological noise model, we find that modest improvements in experimental coherence times to the tens-of-microseconds regime could enable the CZ gate to achieve error rates of approximately $10^{-3}$. These results position the soft \zps as a promising intermediate platform between conventional superconducting qubits and more strongly protected qubits.

To make our numerical modeling tractable, we made several simplifying assumptions.
Our analysis and simulations assume zero-disorder and neglect the $\zeta$ mode, neither of which is realistic. We also model decoherence phenomenologically rather than through a microscopic description of the device environment. These approximations are sufficient to establish the feasibility of the proposed gate protocols, but the resulting fidelities should be interpreted as an upper bound, rather than as a quantitative prediction for a specific experimental device. Moving beyond these approximations is numerically demanding but would be a worthwhile direction for future work.

Looking forward, several important directions are worth pursuing.
First, future theoretical work should quantify how disorder and additional harmonic modes, such as the $\zeta$ mode, affect gate performance. Second, experimental measurements of relaxation and dephasing in higher-energy \zps states would enable more realistic noise models and pulse optimization.  Third, the performance of our gate as devices approach the hard-\zps regime is unclear.  Recent work by \citet{kolesnikow2025} has proposed protected single- and two-qubit gates for hard-\zps devices; incorporating realistic noise into that framework would be a  valuable next step.

\noindent {\em Acknowledgments}: The authors thank Joe Aumentado, Sai Pavan Chitta, Peter Groszkowski, Andras Gyenis, Florent Lecocq, Marco Nicotra, Eyob Sete, Jieqiu Shao, Ray Simmonds, Daniel Slichter, Agustin Di Paolo, and John Teufel for many helpful discussions. JC also thanks Max Block, Nicolas Didier, Andr\'e Melo, Matt Reagor, Eyob Sete, Michael Scheer, Marcus Silva, and Hakan T\"{u}reci for many helpful discussions on circuit QED.
ZL and JC were supported by the National Science Foundation under CAREER Award No. ECCS-2240129. EW was supported by the National Science Foundation Graduate Research Fellowship under Grant No. 2040434.\\

\noindent {\em AI tools statement:} ChatGPT and Claude were used during the later stages of the project to assist with coding and proofreading. The authors reviewed all AI-generated code and text before use.

\appendix

\section{Single-Qubit Simulation Details}\label{app:1Qgate_simulation}
The software used to generate the figures in this manuscript is available in our open-source GitHub repository \cite{zp2q_repo}. Note that we use the ``scQubits'' Python package to construct the single- and two-qubit Hamiltonians \cite{Groszkowski2021scqubitspython,chitta2022computer}. This appendix outlines the numerical techniques used in the single-qubit simulations. For more details, please refer to the source code.

\subsection{Noiseless simulation of single-qubit gates}
To simulate the \zps qubit, we numerically diagonalize the Hamiltonian in \cref{eq:zp Hamiltonian} and keep the lowest $N_{\rm max} +1$ energy levels. 
Before diagonalization, we construct operators for the $\theta$ mode in the charge basis and for the $\phi$ mode in the phase basis with $\phi \in \left[-6\pi, 6\pi \right]$. We set the dimensions of these bases to 181 and 300, respectively. Numerical simulations with larger truncations show that the qubit frequency (the eigenvalue of $\ket{2}$) converges to within $\sim 10^{-8}$ GHz when either basis is reduced by 5. For the highest retained state ($N_{\rm max}=299$), the corresponding maximum change is $\sim 10^{-5}$ GHz. In the diagonal basis, the \zps Hamiltonian is
\begin{equation}
H_{0\text{-}\pi} = \sum_{p=0}^{N_{\rm max}} \omega_{0 \text{-} p} \ket{p} \bra{p},
\end{equation}
where $\ket{p}$ is an eigenvector with eigenvalue $\hbar \omega_{0 \text{-} p}$ (with $\hbar=1$). To add the drive, we express the operator $n_{\rm op}$ with ${\rm op} \in \{\theta, \phi\}$ in the diagonal basis of the \zps Hamiltonian as $n_{\rm op} = \sum_{i,j=0}^{N_{\rm max}} n^{i \text{-} j}_{\rm op} \ket{i} \bra{j}$, where $n^{i \text{-} j}_{\rm op} = \bra{i} n_{\rm op} \ket{j}$ and $\ket{i}, \ket{j} \in \{\ket{p}\}$.

Leakage from the computational subspace into higher excited states is a known issue for the \zps qubit~\cite{PremkumarAPS2023,premkumar2023hamiltonian}. We therefore verify that our finite-level truncation faithfully approximates the dynamics of a converged larger model. We first increase the truncation size $n$ in increments of 100 and calculate the corresponding gate fidelity $F_n$. We assess convergence from the change in infidelity between successive truncations by requiring
$|(1-F_n)-(1-F_{n-100})|\leq (1-F_n)/10,$
such that the truncation-induced change is at least one order of magnitude smaller than the infidelity itself. For example, if $1-F_n=10^{-3}$, we require the change relative to the calculation with $n-100$ states to be no greater than $10^{-4}$. The largest model, with $N_{\rm max}=1000$ states, satisfies this convergence criterion and serves as the reference. We then compare calculations with smaller truncations against the $N_{\rm max}=1000$ result to identify the smallest model that reproduces the reference infidelity to the same level of accuracy. Using this procedure, we find that $N_{\rm max}=300$ states is sufficient for the single-qubit gate simulations.

We also control integration error using QuTiP’s propagator routine. The time-step convergence criterion is similar to that for truncation. We chose the maximum time step so that increasing it by approximately a factor of three changed the infidelity by less than one tenth of the infidelity itself. Thus we used a maximum step size of $\Delta t_{\rm max}=3\times10^{-4}$ ns for $n_\theta$ drives and $1\times10^{-3}$ ns for $n_\phi$ drives in our simulations.

\subsection{Model reduction for noiseless pulse optimization}
\label{app:sec:model_reduce_1q}
 Modeling the \zps requires a large Hilbert space, making simulations slow. Optimizing pulse parameters requires solving the dynamics many times. To balance speed and accuracy, we construct a reduced model for pulse optimization in the \zps qubit. The reduced model includes only states accessible from the logical subspace under the chosen drive.

We build the reduced model by first including the logical states $\ket{0}$ and $\ket{2}$, and the intended intermediate state ($\ket{7}$ for $n_\theta$, $\ket{9}$ for $n_\phi$). 
The drive operator $n_{\rm op}$ with ${\rm op} \in \{\theta, \phi\}$ couples the states in the current reduced model to other states in the Hilbert space. 
Next we compute matrix elements $n_{\rm op}^{i\text{-}j}=\bra{i}n_{\rm op}\ket{j}$ for all $\ket{i}$ in the reduced model and $\ket{j}$ outside it. Any state $\ket{j}$ with $|n_{\rm op}^{i\text{-}j}|\geq n_{\rm thresh}$ is absorbed into the model. Repeating this procedure over the lowest 300 states yields a stable reduced model once no further states are added. We call the resulting set the selected states. For thresholds $n_{\rm thresh}\in[10^{-8},0.2]$, the same reduced models emerge, containing 157 states for $n_\theta$ and 160 for $n_\phi$.

Even with a reduced Hilbert space, simulations remain costly. We gain further efficiency by evolving only the logical states instead of the full propagator. By default, QuTiP v4.7 evolves all basis states, which is unnecessary for gate fidelity optimization. Restricting to logical states greatly speeds up simulations, allowing a 100-ns $X$ gate in a $\sim$200-state space to be optimized in about 1 hour using 30 AMD EPYC 7713 cores.

\subsection{Single-qubit pulse optimization}\label{app:sub_sec_pluse_opt}
\begin{figure*}[!htbp]     
\includegraphics[width=\textwidth]{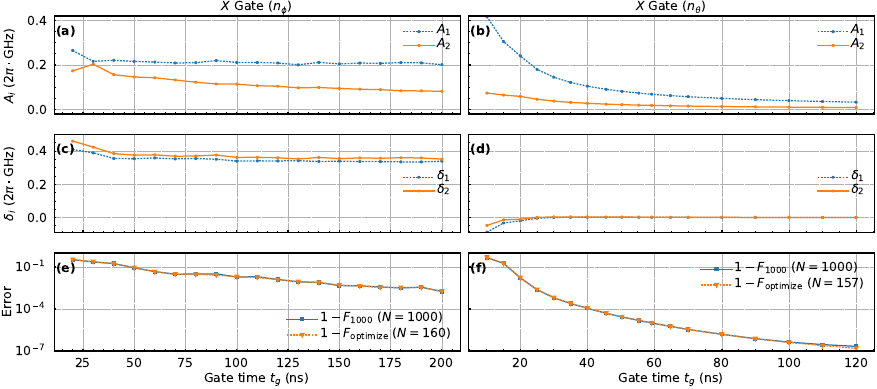}
\centering
\caption {
Optimized pulse parameters and gate infidelities for an $X$ gate driven through $n_{\rm op}$, with ${\rm op}\in\{\phi,\theta\}$. For each gate time $t_g$, panels (a,b) show the optimized amplitudes $(A_1,A_2)$, panels (c,d) show the optimized detunings $(\delta_1,\delta_2)$, and panels (e,f) show the corresponding infidelities. 
The pulses are optimized in a closed-system truncated model with $N_{\rm optimize}=160$ levels for the $n_\phi$ drive and $N_{\rm optimize}=157$ levels for the $n_\theta$ drive. The resulting fidelity is $F_{\rm optimize}$. 
The same pulses are then tested in a closed-system 1000-level model, giving $F_{1000}$.
}
\label{fig:params_1q}
\end{figure*}
To construct an $X$ gate, we use an effective $\Lambda$ system driven on both legs, which unavoidably introduces some leakage. The drive Hamiltonian in \cref{eq:zp_H_drive}, with Gaussian envelopes defined in \cref{eq:gauss_envelope}, is specified by five parameters: gate time $t_g$, amplitudes $A_{1,2}$, and detunings $\delta_{1,2}$. With $t_g$ fixed (tolerance $\pm0.1$ ns), only four parameters are optimized.

Given the pulse parameters, we numerically solve the Schr\"odinger equation,
\begin{equation}
i \frac{\partial V(t)}{\partial t} = \left(H_{0\text{-}\pi} + H_{\rm drive}\right) V(t),
\end{equation}
with $V(0)=I$, to obtain the evolution operator $V(t)$.
 Gate quality is measured by the average fidelity between the target unitary $U_{\rm targ}$ ($=X$) and the simulated propagator $V$ in the logical subspace $\{\ket{0},\ket{2}\}$:
\begin{equation}
F(U_{\rm targ},V)=
\frac{
\operatorname{Tr}(V^\dagger V) +
\big| \operatorname{Tr}(U_{\rm targ}^\dagger V) \big|^2
}
{d(d+1)}\,,
\label{eq:gate_fidelity}
\end{equation}
with $d = 2$. For improved numerical behavior, we optimize over the quantity $\log_{10}(1 - F)$. Due to leakage, $V$ is generally not unitary, although it is close to unitary for high-fidelity operations.

The optimization is constrained by $A_i/2\pi \in (0,\ 500)\ \text{MHz}$ and $\delta_i/2\pi \in (-500,\ 500)\ \text{MHz}$.  We first optimize in the reduced model from \cref{app:sec:model_reduce_1q}, then re-run the search using the result as an initial guess to avoid local minima. The optimization uses ``differential evolution'', a gradient-free stochastic method~\cite{storn1997differential} that efficiently explores large parameter spaces in parallel. Finally, we validate the optimized pulse by simulating the system in a larger Hilbert space spanned by the lowest 1000 eigenstates.

After each optimization, we record the infidelity and use the result as the initial guess for the next iteration to improve convergence. This process is repeated for all gate times. The optimized pulse parameters obtained for the previous gate time are used as the initial guess for the optimization at the next gate time. This warm-start strategy significantly accelerates the optimization.

In \cref{fig:params_1q}, we show the optimized pulse parameters and gate infidelity for the $X$ gate with $n_\theta$ and $n_\phi$ drives. Panels (a)–(b) display the drive amplitudes. The amplitude ratio $A_1/A_2$ is largely set by the ratio of charge matrix elements $n_{\rm op}^{2\text{-}i}/n_{\rm op}^{0\text{-}i}$.  In an ideal three-level $\Lambda$ system these ratios coincide, giving the fastest transfer. Due to transitions nearby in a large Hilbert space, we need to optimize $A_1$ and $A_2$ independently.
For the optimized pulse parameters shown in \cref{fig:params_1q}, the amplitude ratio  $A_1/A_2$ agrees well with the matrix element ratio, $n_{\rm op}^{2\text{-}i} / n_{\rm op}^{0\text{-}i}$ for $n_\theta$ driving. This agreement does not hold for $n_\phi$ driving. However, the ratio of the drive amplitudes converges closely to the ratio of the charge matrix elements as the gate time increases beyond 200 ns for $n_\phi$ driving. This is also the case for the $\sim$800 ns $X$ gate, achieved via driving $n_\phi$, presented in \cref{sec:x_gate_reproduce}. Also visible in (b) is the decrease in amplitudes with gate time, which correlates with reduced gate error in (f). Panels (c)–(d) show the detunings, indicating off-resonant vs. resonant driving for $n_\phi$ and $n_\theta$. Panels (e)–(f) show gate infidelities for the full model with 1000 states ($1-F_{n_{\rm op}}$) and two reduced models.

To get an idea of the order of magnitude of the error incurred by truncation,  we also compute the relative gate-error difference $|F_{1000}-F_{\rm optimize}|/(1-F_{1000})$ in \cref{fig:params_1q}(e)–(f). 
For $n_\theta$ driving, the average and maximum differences are $1.16\times10^{-2}$ and $5.56\times10^{-2}$, respectively. For $n_\phi$ driving, they are $4.59\times10^{-2}$ and $1.47\times10^{-1}$. 

\subsection{Noisy simulation of single-qubit gates}\label{app:sub_sec_noisy_1Q}
We simulate noise using the master equation $\dot\rho=\mathcal{L}[\rho]$, whose solution is $\rho(t)=e^{\mathcal{L}t}[\rho(0)]$, with $e^{\mathcal{L}t}$ the propagator from $\rho(0)$ to $\rho(t)$. The noise model includes energy relaxation ($T_1$) and pure dephasing ($T_\phi$), with equal characteristic times $T_1=T_\phi=T$. We consider $T\in\{3,30,170\}\,\mu\mathrm{s}$. These values are chosen to represent noise regimes corresponding approximately to gate fidelities of $\sim99\%$, $\sim99.9\%$, and above $99.9\%$, respectively.

To compute gate fidelities, we fix a target gate time and use the noiselessly optimized pulse parameters (drive frequencies and amplitudes). We then integrate the master equation to obtain the superoperator propagator $\tilde{\mathcal{X}}=e^{\mathcal{L}t}$, with Liouvillian $\mathcal{L}$. At the final time, $\tilde{\mathcal{X}}$ is converted into Kraus operators $V_k$, defining the noisy channel $V(\rho)=\sum_k V_k \rho V_k^\dagger$.
The average channel fidelity relative to the target $X$ gate is then
\begin{equation}
F(U_{\rm targ},V)
\!=\!
\frac{1}{d(d+1)}
\displaystyle
\sum_k \left[
\operatorname{Tr}(V_k^\dagger V_k) +
\big| \operatorname{Tr}(U_{\rm targ}^\dagger V_k)
\big|^2 \right],
\label{eq:channel_fidelity}
\end{equation}
with $U_{\rm targ}=X$ and $d=2$. Because energy-relaxation rates scale with charge matrix elements, they vanish when the matrix element is negligible. 
Because open-system simulations are computationally more demanding, we employ a smaller Hilbert-space truncation than in the corresponding closed-system simulations.
For noisy single-qubit simulations, we evolve the optimized pulse in a reduced subspace of $N$ selected states, with $N=160$ for the $n_\phi$ drive and $N=157$ for the $n_\theta$ drive.
The resulting superoperator is then used to compute the noisy channel fidelity relative to the ideal $X$ gate.

\section{Comparison to Experimental Single-Qubit Gate Performance}
\label{sec:x_gate_reproduce}

To validate our methods, we simulate a gate similar to the $X$ gate demonstrated in Ref.~\cite{gyenis2021}. It used a Gaussian pulse with detuning $\delta\approx-4$ MHz and $\sigma=200$ ns, so $t_g\approx800$ ns. That gate achieved a fidelity $\leq90\%$ \cite{andras_xgate}.

We already use the same device parameters (such as $E_J$, $E_C$, etc.) as in the experiment. To replicate their resonator-pin drive, we model the drive operator as $n_{\rm op} \propto b_\phi n_\phi + b_\theta n_\theta$, where the coupling rates are $b_\phi = 0.27$ and $b_\theta = 6.6 \times 10^{-3}$~\cite{gyenis2021}. With this model, we numerically optimize a pulse and find a gate with $t_{\rm g} \approx 830$ ns, which is of the same order of magnitude as the gate duration in Ref.~\cite{gyenis2021}. This pulse was optimized on the closed system model, not the open system model. It would likely be improved by directly optimizing the open system model.

\begin{figure}[!hbtp] 
\includegraphics[width=\columnwidth]{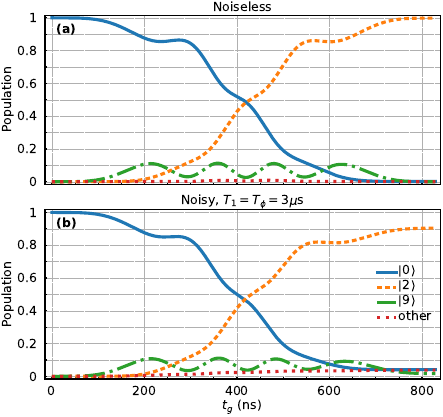}
\centering
\caption {
Population transfer during an $X$ gate with $t_{\rm g}=828.76$ ns, driven by $n_{\rm op} \propto b_\theta n_\theta + b_\phi n_\phi$. (a) Noiseless simulation with fidelity $F=99.85\%$. (b) Noisy simulation with $T_1=T_\phi=3~\mu$s, reducing fidelity to $F=85.72\%$. The experimental drive coefficients were $b_\theta=6.6\times10^{-3}$ and $b_\phi=0.27$. For consistency with the $n_\phi$ and $n_\theta$ driving in the main text, we normalize these coefficients while preserving their ratio, giving $n_{\rm op}=0.976\, n_\phi+0.024\,n_\theta$ in the simulations.
}\label{fig:exp_gate_population}
\end{figure}

\Cref{fig:exp_gate_population} shows the noiseless population transfer for this gate. With parameters $A_1=13.56$ MHz, $A_2=34.96$ MHz, $\delta_1=-3.03$ MHz, $\delta_2=-3.18$ MHz, and $t_g=828.76$ ns, we obtain a noiseless fidelity of $F=99.85\%$. Including noise with $T_1=T_\phi=3\ \mu$s reduces the fidelity to $F=85.72\%$, which is comparable to the experimental value. This demonstrates that our model captures key features of the experimental gate.

There is an apparent discrepancy between the gate times studied here and those in the main text. The gate in this appendix is significantly slower, and its closest counterpart in the main text is the $n_\phi$-driven gate.

The main reason is the much larger effective Rabi rates achievable in our protocol. Ref.~\cite{gyenis2021} demonstrated Raman control with effective Rabi frequencies of only a few MHz. However, this limitation was not intrinsic to the soft 0-$\pi$ qubit itself. Rather, as noted in Ref.~\cite{gyenis2021}, the control tones were applied through the readout cavity, whose strong filtering substantially reduced the drive power reaching the qubit. In contrast, our simulations assume direct microwave control and exploit intermediate states with relatively large transition matrix elements. The optimized parameters in \cref{tab:xgate_parameters} yield effective single-photon Rabi rates up to $\Omega/2\pi \approx 60~\mathrm{MHz}$, approximately an order of magnitude larger than those reported experimentally. These larger effective Rabi rates are a major factor behind the significantly shorter gate times found in the main text.

A second difference is the drive operator. In this appendix, the drive has the form $b_\phi n_\phi + b_\theta n_\theta$, whereas the main text studies gates driven by $n_\phi$ or $n_\theta$ alone. Simulations not shown here indicate that driving the combined operator $b_\phi n_\phi + b_\theta n_\theta$ produces more leakage than driving either operator
alone. These additional leakage pathways reduce the gate fidelity and likely contribute to the slower gate in Ref.~\cite{gyenis2021}.

Finally, there are likely differences associated with noise and leakage modeling. Our model captures key experimental features, but it is not a complete device-level model. In particular, our model does not include a $\zeta$ mode, and higher excited states may decohere more strongly in the experiment than assumed in our simulations. The main-text $n_\phi$ gate also reaches a larger intermediate-state population than the gate in this appendix, with $P_9\sim 0.22$ compared with $P_9\sim 0.12$. Reducing this population is possible, but generally requires longer gate times. For comparison, the simulation in Appendix G of Ref.~\cite{gyenis2021} shows an intermediate-state population of approximately $5.5\%$ in Fig.~16(e), likely due to the use of an effective three-level Hilbert space in that calculation.

After submission, we became aware of the thesis by \citet{premkumar2023hamiltonian}. We do not directly compare our model with the results in Fig.~4.9 of the thesis because the simulations differ in several important respects. In particular, the simulated device uses parameters that differ substantially from those of Ref.~\cite{gyenis2021}, the gate employs a flat-top Gaussian pulse rather than the pulse considered here, and the two Raman tones are constrained to have a common detuning from their respective transitions. By contrast, we optimize the two drive detunings independently.


\section{Quantization of two coupled \texorpdfstring{\zp}{zero-pi} qubits }\label{app:quantize}

In \cref{app:coupling_scheme_theta} we first quantize the circuit in \cref{fig:circuit_2qubits}(a), which produces an  $ n_{\theta1} n_{\theta 2} $ coupling. We then quantize the circuit in \cref{fig:circuit_2qubits}(b), which produces an  $ n_{\theta1}n_{\theta2} - n_{\phi1}n_{\phi2} $  coupling.

\subsection{Pure coupling scheme: \texorpdfstring{$ n_{\theta1} n_{\theta 2} $} {theta} }\label{app:coupling_scheme_theta}

Our goal is to quantize the 10-node circuit in \cref{fig:circuit_2qubits} (a). The derivation below assumes equal capacitances for both \zps circuits. If we instead allowed different capacitances for the two qubits, the coupling strength and $\theta$ mode capacitance would be slightly modified. Crucially, this does not change the circuit quantization procedure, variable transformation, or the form of the effective Hamiltonian.

Let $T_{0\text{-}\pi}^{(i)}$ and $U_{0\text{-}\pi}^{(i)}$ denote the kinetic (capacitive) and potential terms of the Lagrangian for the $i^\text{th}$ \zps qubit, respectively, where $i \in \{1, 2\}$. Let $T_{\rm int}$ represent the interaction between the two qubits. Then, the full system Lagrangian is
\begin{align}\label{eq:full_sys_lagrangian}
\mathcal{L}
=& \sum_{i=1,2}{(T_{0\text{-}\pi}^{(i)} - U_{0\text{-}\pi}^{(i)} )} + T_{\rm int} \, .
\end{align}
Here, the superscript $i\in\{1,2\}$ labels the qubit, and $T_{\rm int}$ represents the interaction between the two \zps qubits.  $T_{0\text{-}\pi}^{(i)}$ and $U_{0\text{-}\pi}^{(i)} $ are the kinetic and potential parts of the single-qubit Lagrangian, respectively. The kinetic terms for the two \zps qubits are
\begin{widetext}
\begin{align}
T_{0\text{-}\pi}^{(1)}
&= (C+C_{J})\dot{\theta}_1^2 +  C_J\dot{\phi}_1^2 + C\dot{\zeta}_1^2 + \half C_0 \sum_{i=1}^{4}{\dot{\varphi}_i^2} \nonumber\\
& =  ( C+C_J+\half C_0  ) \dot{\theta}_1^2 +  (C_J+\half C_0) \dot{\phi}_1^2 + ( C+\half C_0 ) \dot{\zeta}_1^2 +\half C_0  \dot{\Sigma}_1^2,  \\
T_{0\text{-}\pi}^{(2)}
&= (C+C_{J})\dot{\theta}_2^2 +  C_J\dot{\phi}_2^2 + C\dot{\zeta}_2^2 + \half C_0  \sum_{i=5}^{8}{\dot{\varphi}_i^2} \nonumber \\ 
&= ( C+C_J+\half C_0 ) \dot{\theta}_2^2 +  (C_J+\half C_0) \dot{\phi}_2^2 + ( C+\half  C_0 ) \dot{\zeta}_2^2 +\half C_0 \dot{\Sigma}_2^2 \, .
\end{align}
In these expressions, the top line has the node variables and the bottom line is re-expressed with respect to our variable transformation in \cref{eq:zp_transform}. The additional terms relative to the single \zps circuit, proportional to $C_0$, arise from the ground capacitors. The coupling term is more complicated and can be written as
\begin{align}
T_{\rm int}
&=\frac{C_{c}}{2} 
[(\dot{\varphi }_1-\dot{\varphi }_{10}){}^2 + 
(\dot{\varphi }_4-\dot{\varphi }_{10}){}^2 + 
(\dot{\varphi }_5-\dot{\varphi }_{10}){}^2 +     
(\dot{\varphi }_8-\dot{\varphi }_{10}){}^2 \nonumber
\\
&\quad +(\dot{\varphi }_2-\dot{\varphi }_9){}^2 + 
(\dot{\varphi }_3-\dot{\varphi }_9){}^2 + 
(\dot{\varphi }_6-\dot{\varphi }_9){}^2 + 
(\dot{\varphi }_7-\dot{\varphi }_9){}^2]  \nonumber
\\
&= \frac{C_{c}}{2} \Big [ \sum_{i=1}^{8}{\dot{\varphi}_i^2} 
+ 4(\dot{\varphi}_9^2 + \dot{\varphi}_{10}^2)
-2\dot{\varphi}_9 (\dot{\varphi}_2 + \dot{\varphi}_3 + \dot{\varphi}_6 + \dot{\varphi}_7)
-2\dot{\varphi}_{10} (\dot{\varphi}_1 + \dot{\varphi}_4 + \dot{\varphi}_5 + \dot{\varphi}_8) \Big ]   \nonumber
 \\
&=\frac{C_c}{2}\sum_{i=1}^{2} [\dot{\phi}_i^2 + \dot{\zeta}_i^2  
+(\dot{\varphi }_+ - \dot{\Sigma }_i)^2 
+(\dot{\varphi }_- - \dot{\theta }_i)^2 ] \, .
\end{align}
The final line is written using the variable transformation in \cref{eq:zp_transform} and the sum and difference variables
$\varphi_{\pm}=\varphi_9\pm\varphi_{10}$. Finally, the potential term is 
\begin{equation}
U_{0\text{-}\pi}^{(i)}
=-2E_{Ji} \cos{\theta_i} \cos{(\phi_i-\phiext{}_{,i}/2)} + E_{Li}\phi_i^2 + E_{Li} \zeta_i^2 .
\end{equation}
After doing the Legendre transform, the vector of conjugate momenta $Q_\mu=\partial\mathcal{L}/\partial\dot{\mu}\ (\mu=\theta_j,\phi_j,\zeta_j,\Sigma_j,+,-;\ j=1,2)$ is
\begin{align} \label{eq:capacitance_mat1}
\left(\begin{array}{c}
Q_{\theta 1} \\
Q_{\Sigma 1} \\
Q_{\phi 1} \\
Q_{\zeta 1} \\
Q_{\theta 2} \\
Q_{\Sigma 2} \\
Q_{\phi 2} \\
Q_{\zeta 2} \\
Q_{+}\\
Q_{-}
\end{array}\right)=\left(\begin{array}{cccccccccc}
C_\theta & 0 & 0  & 0 & 0  & 0 & 0  & 0 & 0  & -C_c \\
0 & C_\Sigma & 0  & 0 & 0  & 0 & 0  & 0 & -C_c  & 0 \\
0 & 0 & C_\phi  & 0 & 0  & 0 & 0  & 0 & 0  & 0 \\
0 & 0 & 0  & C_\zeta & 0  & 0 & 0  & 0 & 0  & 0 \\
0 & 0 & 0  & 0 & C_\theta  & 0 & 0  & 0 & 0  & -C_c \\
0 & 0 & 0  & 0 & 0  & C_\Sigma & 0  & 0 & -C_c  & 0 \\
0 & 0 & 0  & 0 & 0  & 0 & C_\phi  & 0 & 0  & 0 \\
0 & 0 & 0  & 0 & 0  & 0 & 0  & C_\zeta & 0  & 0 \\
0 & -C_c & 0  & 0 & 0  & -C_c & 0  & 0 & 2C_c  & 0 \\
-C_c & 0 & 0  & 0 & -C_c  & 0 & 0  & 0 & 0  & 2C_c 
\end{array}\right)\left(\begin{array}{c}
\dot{\theta}_1 \\
\dot{\Sigma}_1 \\
\dot{\phi}_1 \\
\dot{\zeta}_1 \\
\dot{\theta}_2 \\
\dot{\Sigma}_2 \\
\dot{\phi}_2 \\
\dot{\zeta}_2 \\
\dot{\varphi}_+\\
\dot{\varphi}_-
\end{array}\right) \, .
\end{align}
In the above equation, the capacitances $C_{\mu}$ of the \zps qubit modes  for $\mu \in \{\theta,\ \phi,\ \zeta,\ \Sigma\}$ are specified by 
\begin{align} \label{eq:capacitance_transform}
\left(\begin{array}{c}
C_{\theta} \\
C_{\phi} \\
C_{\zeta} \\
C_{\Sigma} \\
\end{array}\right)=\left(\begin{array}{cccc}
2 & 2 & 1 & 1  \\
0 & 2 & 1 & 1  \\
2 & 0 & 1 & 1  \\
0 & 0 & 1 & 1  \\
\end{array}\right)\left(\begin{array}{c}
C \\
C_J \\
C_c \\
C_0 \\
\end{array}\right) \, 
\end{align}
where $C_0$ is the capacitance between the node and ground. With a less intuitive choice of intermediate variables, 
\begin{equation}
    \varphi_+' = \varphi_+-(\Sigma_1+\Sigma_2)/2\quad \text{and}\quad \varphi_-' = \varphi_--(\theta_1+\theta_2)/2,
\end{equation} 
the coupling between the $\theta$ and $\Sigma$ modes of the two qubits can be seen directly:
\begin{align} \label{eq:capacitance_mat}
\left(\begin{array}{c}
Q_{\theta 1} \\
Q_{\Sigma 1} \\
Q_{\phi 1} \\
Q_{\zeta 1} \\
Q_{\theta 2} \\
Q_{\Sigma 2} \\
Q_{\phi 2} \\
Q_{\zeta 2} \\
Q_{+'}\\
Q_{-'}
\end{array}\right)=\left(\begin{array}{cccccccccc}
C_\theta-\frac{C_c}{2} & 0 & 0  & 0 & -\frac{C_c}{2}  & 0 & 0  & 0 & 0  & 0 \\
0 & C_\Sigma -\frac{C_c}{2}& 0  & 0 & 0  & -\frac{C_c}{2} & 0  & 0 & 0  & 0 \\
0 & 0 & C_\phi  & 0 & 0  & 0 & 0  & 0 & 0  & 0 \\
0 & 0 & 0  & C_\zeta & 0  & 0 & 0  & 0 & 0  & 0 \\
-\frac{C_c}{2} & 0 & 0  & 0 & C_\theta -\frac{C_c}{2}  & 0 & 0  & 0 & 0  & 0 \\
0 & -\frac{C_c}{2} & 0  & 0 & 0  & C_\Sigma -\frac{C_c}{2}& 0  & 0 & 0  & 0 \\
0 & 0 & 0  & 0 & 0  & 0 & C_\phi  & 0 & 0  & 0 \\
0 & 0 & 0  & 0 & 0  & 0 & 0  & C_\zeta & 0  & 0 \\
0 & 0 & 0  & 0 & 0  & 0 & 0  & 0 & 2C_c  & 0 \\
0 & 0 & 0  & 0 & 0  & 0 & 0  & 0 & 0  & 2C_c 
\end{array}\right)\left(\begin{array}{c}
\dot{\theta}_1 \\
\dot{\Sigma}_1 \\
\dot{\phi}_1 \\
\dot{\zeta}_1 \\
\dot{\theta}_2 \\
\dot{\Sigma}_2 \\
\dot{\phi}_2 \\
\dot{\zeta}_2 \\
\dot{\varphi}_+'\\
\dot{\varphi}_-'
\end{array}\right) \, .
\end{align}

Calculating the inverse of the capacitance matrix in \cref{eq:capacitance_mat} gives the total quantized Hamiltonian  $H=\frac{1}{2}Q^TC^{-1}Q + U$, which can be decomposed as
\begin{align}\label{eq:H_tot_apxA1}
    H &= H_{0\text{-}\pi}^{(1)} + H_{0\text{-}\pi}^{(2)} + H_{\rm int} + H_{\rm coupler}, 
\end{align}
where $H_{\rm int}$ represents the interaction between the two qubit circuits and the coupling circuit, while $H_{\rm coupler}$  is the coupler self-Hamiltonian.
The Hamiltonian of the individual \zps qubits is
\begin{align}\label{eq:zp_h_app_i}
    H_{0\text{-}\pi}^{(i)} &=  4\tilde E_C^{\theta} {n}_{\theta i}^2 + 4\tilde E_C^{\phi} {n}_{\phi i}^2  + 4\tilde E_C^{\zeta} {n}_{\zeta i}^2 + 4\tilde E_C^{\Sigma} {n}_{\Sigma i}^2 + 
    E_{Li}{\phi}_i^2 + E_{Li}{\zeta}_i^2 - 2E_{Ji}\cos{{\theta}_i}\cos \Big ({\phi}_i - \frac{\phiext}{2}\Big) , 
\end{align}
where the associated energies are
\begin{subequations}
\begin{align}\label{eq:renormalized_ec}
    \tilde E_{C}^{\theta} &= \frac{e^2(2C_\theta-C_c)}{4C_{\theta}(C_{\theta} - C_c)},
   & \tilde E_{C}^{\phi} &= \frac{e^2}{2C_\phi},
   \\
    \tilde E_{C}^{\Sigma} &= \frac{e^2(2C_\Sigma-C_c)}{4C_{\Sigma}(C_{\Sigma} - C_c)} ,
   & \tilde E_{C}^{\zeta} &= \frac{e^2}{2C_\zeta} \, .
\end{align}
\end{subequations}
The two remaining parts of \cref{eq:H_tot_apxA1} are
\begin{subequations}
\begin{align}
    H_{\rm coupler} = 4E_C^- n_m^2 + 4E_C^+ n_p^2 , 
    \quad {\&}\quad
    H_{\rm int} = g_{\theta} n_{\theta_1}n_{\theta_2} + g_{\Sigma} n_{\Sigma_1}n_{\Sigma_2},
\end{align}
\end{subequations}
with the coefficients 
\begin{subequations}\label{eq:coupling_rates_app}
\begin{align}
    g_{\theta} &=\frac{2e^2C_c}{C_{\theta} (C_{\theta} - C_c)} \approx \frac{e^2C_c}{2C^2}, 
    &E_{C}^{+} &=\frac{e^2}{4C_c},
    \\
    g_{\Sigma} &= \frac{2e^2C_c}{C_{\Sigma}(C_{\Sigma} - C_c)} = \frac{2e^2C_c}{C_0(C_0+C_c)}, &E_{C}^{-} &= \frac{e^2}{4C_c}.
\end{align}
\end{subequations}
The approximate expressions are valid in the limit of $C\gg C_c,C_J,C_0$. 
In the above expressions, the modes $\zeta_1$ and $\zeta_2$ are decoupled from the effective modes of the \zps qubit and can thus be considered separately. The modes $\Sigma_1$, $\Sigma_2$, $\varphi_+$, and $\varphi_-$ do not have corresponding potential energy terms and are considered free modes. Because they are decoupled from the $\theta, \phi,$ and $\zeta$ degrees of freedom, the free modes may be ignored without affecting the system dynamics. Thus the final Hamiltonian can be written as 
\begin{subequations}
\begin{align}
    H_{\rm tot} &= H_{0\text{-}\pi}^{(1)} + H_{0\text{-}\pi}^{(2)} + H_{\rm int}, 
    \\
    H_{0\text{-}\pi}^{(i)} &=  4\tilde E_C^{\theta} n_{\theta i}^2 + 4\tilde E_C^{\phi} n_{\phi i}^2   +   E_{Li}\phi_i^2 - 2E_{Ji}\cos{\theta_i}\cos\Big (\phi_i - \frac{\phiext}{2}\Big) , 
    \\
    H_{\rm int} &= g_{\theta} n_{\theta_1}n_{\theta_2}.
\end{align}
\end{subequations}

\subsection{ Mixed coupling scheme: \texorpdfstring{$ n_{\theta1}n_{\theta2} - n_{\phi1}n_{\phi2} $}{theta + phi}}\label{app:coupling_scheme_mix}
Now we quantize the circuit in \cref{fig:circuit_2qubits}(b). We use the variable transformation in \cref{eq:zp_transform} on both \zps qubits.  The Lagrangian is expressed, as before, in \cref{eq:full_sys_lagrangian}. Recall that the superscript $i\in\{1,2\}$ labels the qubit; $T_{\rm int}$ is the qubit–qubit interaction; $T_{0\text{-}\pi}^{(i)}$ and $U_{0\text{-}\pi}^{(i)}$ are the kinetic (capacitive) and potential parts of the Lagrangian, respectively. The kinetic part of the Lagrangian for the two \zps qubits can be written compactly as
$T_{0\text{-}\pi}^{(i)}
= ( C+C_J+\frac{C_0}{2} ) \dot{\theta}_i^2 +  (C_J+\frac{C_0}{2}) \dot{\phi}_i^2 + ( C+\frac{C_0}{2} ) \dot{\zeta}_i^2 +\frac{C_0}{2} \dot{\Sigma}_i^2$. 
The coupling term is more complicated and can be written as
\begin{align}
T_{\rm int}&=\frac{C_{c}}{2} 
[(\dot{\varphi }_1-\dot{\varphi }_8){}^2 + 
(\dot{\varphi }_2-\dot{\varphi }_7){}^2     
+ (\dot{\varphi }_3-\dot{\varphi }_6){}^2 + 
(\dot{\varphi }_4-\dot{\varphi }_5){}^2]     \nonumber
\\
&=\frac{C_{c}}{2} 
[(\dot{\theta }_1 - \dot{\theta }_2){}^2 + 
(\dot{\phi }_1 + \dot{\phi }_2){}^2     
+ (\dot{\zeta }_1 + \dot{\zeta }_2){}^2 + 
(\dot{\Sigma }_1 - \dot{\Sigma }_2){}^2].
\end{align}
Finally, the potential term is 
\begin{align}
U_{0\text{-}\pi}^{(i)} = -2E_{Ji} \cos{\theta_i} \cos{(\phi_i-\phiext{}_{,i}/2)} + E_{Li}\phi_i^2 + E_{Li} \zeta_i^2 \, .
\end{align}
After doing the Legendre transform, the vector of conjugate momenta $Q_\mu=\partial\mathcal{L}/\partial\dot{\mu}\ (\mu=\theta_j,\phi_j,\zeta_j,\Sigma_j;\ j=1,2)$ is
\begin{align} \label{eq:capacitance_mat_2}
\left(\begin{array}{c}
Q_{\theta 1} \\
Q_{\Sigma 1} \\
Q_{\phi 1} \\
Q_{\zeta 1} \\
Q_{\theta 2} \\
Q_{\Sigma 2} \\
Q_{\phi 2} \\
Q_{\zeta 2} \\
\end{array}\right)=\left(\begin{array}{cccccccc}
C_\theta & 0 & 0  & 0 & -C_c  & 0 & 0  & 0  \\
0 & C_\Sigma & 0  & 0 & 0  & -C_c & 0  & 0 \\
0 & 0 & C_\phi  & 0 & 0  & 0 &  C_c  & 0  \\
0 & 0 & 0  & C_\zeta & 0  & 0 & 0  & C_c  \\
-C_c & 0 & 0  & 0 & C_\theta  & 0 & 0  & 0  \\
0 & -C_c & 0  & 0 & 0  & C_\Sigma & 0  & 0  \\
0 & 0 &  C_c  & 0 & 0  & 0 & C_\phi  & 0  \\
0 & 0 & 0  &  C_c & 0  & 0 & 0  & C_\zeta  \\
\end{array}\right)\left(\begin{array}{c}
\dot{\theta}_1 \\
\dot{\Sigma}_1 \\
\dot{\phi}_1 \\
\dot{\zeta}_1 \\
\dot{\theta}_2 \\
\dot{\Sigma}_2 \\
\dot{\phi}_2 \\
\dot{\zeta}_2 \\
\end{array}\right).
\end{align}
Here, the capacitances $C_{\mu}$ of the \zps qubit modes $\theta,\ \phi,\ \zeta,\ \Sigma$ are specified by \cref{eq:capacitance_transform}. 

Calculating the inverse of the capacitance matrix in \cref{eq:capacitance_mat_2} gives the total quantized Hamiltonian  $H=\frac{1}{2}Q^T C^{-1} Q + U$, which can be decomposed as
\begin{align} \label{eq:H_tot_apxA2}
    H =& H_{0\text{-}\pi}^{(1)} + H_{0\text{-}\pi}^{(2)} + H_{\rm int}, 
\end{align}
where $H_{\rm int}$ represents the interaction between the two qubit circuits.

The Hamiltonian of the individual \zps qubits is 
\begin{align}
    H_{0\text{-}\pi}^{(i)} =&  4\tilde E_C^{\theta} n_{\theta i}^2 + 4\tilde E_C^{\phi} n_{\phi i}^2  + 4 \tilde E_C^{\zeta} n_{\zeta i}^2 + 4\tilde E_C^{\Sigma} n_{\Sigma i}^2 + 
    E_{Li}\phi_i^2 + E_{Li}\zeta_i^2 - 2E_{Ji}\cos{\theta_i}\cos{\Big(\phi_i - \frac{\phiext}{2}\Big)} , 
\end{align}
where the associated energies are
\begin{subequations}
\begin{align}
    \tilde E_{C}^{\theta} =& \frac{e^2C_\theta}{2(C_{\theta}^2 - C_c^2)},
    &\tilde E_{C}^{\phi} &= \frac{e^2C_\phi}{2(C_{\phi}^2 - C_c^2)} ,
    \\
    \tilde E_{C}^{\Sigma} =& \frac{e^2C_\Sigma}{2(C_{\Sigma}^2 - C_c^2)}
    ,\;\;\;\;
    &\tilde E_{C}^{\zeta} &= \frac{e^2C_\zeta}{2(C_{\zeta}^2 - C_c^2)} 
\end{align}
\end{subequations}
The approximate expressions are valid in the limit of $C\gg C_c,C_J,C_0$. The remaining part of \cref{eq:H_tot_apxA2} is 
\begin{align}\label{eq:coupling_2_Hint}
    H_{\rm int} =& g_{\theta} n_{\theta_1}n_{\theta_2} 
    - g_{\phi} n_{\phi_1}n_{\phi_2} 
    - g_{\zeta} n_{\zeta_1}n_{\zeta_2}
    + g_{\Sigma} n_{\Sigma_1}n_{\Sigma_2},
\end{align}
where the coefficients are given by 
\begin{subequations}\label{eq:coupling_2_rates}
\begin{align}
    g_{\theta} =&\frac{4e^2C_c}{C_{\theta}^2 - C_c^2} \approx \frac{e^2C_c}{C^2},
    & g_{\phi} =&\frac{4e^2C_c}{C_{\phi}^2 - C_c^2} = \frac{4e^2C_c}{(2C_J+C_0+2C_c) (2C_J+C_0)} ,
    \\
    g_{\zeta} =&\frac{4e^2C_c}{C_{\zeta}^2 - C_c^2} \approx \frac{e^2C_c}{C^2},
    &g_{\Sigma} =& \frac{4e^2C_c}{C_{\Sigma}^2 - C_c^2} =  \frac{4e^2C_c}{C_0(C_0+2C_c)}.
\end{align}
\end{subequations}
The approximate expressions are valid in the limit of $C\gg C_c,C_J,C_0$. One important point about \cref{eq:coupling_2_Hint} is that all modes of the two \zps circuits are included in this coupling scheme. This is because the coupling circuit does not address any particular mode of the circuits. From \cref{eq:coupling_2_rates} it is clear that to increase or decrease $g_\theta$ relative to  $g_\phi$, we need to vary $C$, which may contradict moving towards a hard \zps regime. 

Here, the modes $\zeta_1$ and $\zeta_2$ are still decoupled from the effective modes of their respective \zps qubits and can thus be removed from the Hamiltonian. Again, the ``free modes'' $\Sigma_1$ and $\Sigma_2$ do not have corresponding potential energy terms and can be ignored without affecting the system dynamics. Thus the final Hamiltonian can be written as 
\begin{subequations}
\begin{align}
    H =& H_{0\text{-}\pi}^{(1)} + H_{0\text{-}\pi}^{(2)} + H_{\rm int}, \\
    H_{0\text{-}\pi}^{(i)} =&  4\tilde E_C^{\theta} n_{\theta i}^2 + 4\tilde E_C^{\phi} n_{\phi i}^2   +   E_{Li}\phi_i^2 - 2E_{Ji}\cos{\theta_i}\cos\left (\phi_i - \frac{\phiext}{2}\right ) , \\
    H_{\rm int} =& g_{\theta} n_{\theta_1}n_{\theta_2}
    - g_{\phi} n_{\phi_1}n_{\phi_2}.
\end{align}
\end{subequations}
\end{widetext}

\section{CNOT gate via a two-tone Raman pulse}\label{sec:cnot}

In this appendix, we introduce a two-qubit CNOT gate for the \zps qubit. The gate uses the $n_{\theta 1}n_{\theta 2}$ coupling derived in \cref{sec:coupling} to realize an effective $\Lambda$ system involving a non-computational state. While its operating principle resembles that of the single-qubit gates in \cref{sec:1q}, it is significantly slower than both the single-qubit gates and the CZ gate of \cref{sec:cz}. A potential advantage of this scheme is its compatibility with stimulated Raman adiabatic passage (STIRAP), which can suppress occupation of the intermediate state and thereby mitigate leakage.

The CNOT we construct is controlled on qubit two
\begin{equation}
\hspace{2pt}
\Qcircuit @C=1em @R=1.7em {
\lstick{1}& \targ{1} & \qw \\
\lstick{2}& \ctrl{-1} & \qw   
}
\hspace{3pt} \raisebox{-10pt}{,}
\end{equation}
with a corresponding unitary
\begin{align} \label{eq:cnot01}
 {\rm CNOT}_{21} = I \otimes \op{0_L}{0_L} + X \otimes \op{1_L}{1_L} \, .
\end{align}
The subscript of $ {\rm CNOT}_{21}$ indicates the control and target qubits.
For comparison, ${\rm CNOT}_{12} = \op{0_L}{0_L} \otimes I + \op{1_L}{1_L} \otimes X$ places the control on qubit 1. ${\rm CNOT}_{12}$ can be implemented by driving $n_{\theta 2}$.

\subsection{Driven Hamiltonian }\label{sec_sub:cnot_drive}
\begin{figure}[!htbp] 
    \includegraphics[width=1\columnwidth]{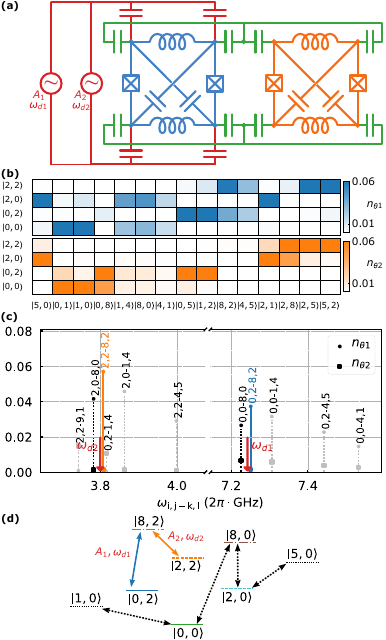}
\centering
\caption {   CNOT gate implemented by driving $n_{\theta 1}$.
(a) Circuit diagram. The coupling and drive circuits are depicted in green and red, respectively. 
(b) Charge matrix elements of $n_{\theta 1}$ and $n_{\theta 2} $ between two-qubit logical states and some non-logical states. Note that the matrix elements used in the CNOT gate correspond to the $n_{\theta 1}$ matrix elements for the $\ket{2,2}$-$\ket{8,2}$ and $\ket{0,2}$-$\ket{8,2}$ transitions.
(c) Charge matrix elements $n_{\theta 1}$ and $n_{\theta 2}$ versus transition frequency $\omega_{i,j \text{-} k,l}$ for a transition $\ket{i,j}$-$\ket{k,l}$. The two Raman transitions are depicted in blue and orange, while the two dominant leakage transitions are depicted in green and red.  
(d) Energy level diagram. The Raman and leakage transitions are plotted using the same color scheme as in panel (c). 
}\label{fig:cnot_ntheta}
\end{figure}

The driven coupled circuit from \cref{sec:coupling} is shown in \cref{fig:cnot_ntheta}(a), with total Hamiltonian
\begin{align}\label{eq:H_total_and_drive_cnot}
    H_{\rm tot} = H_{0\text{-}\pi}^{(1)} + H_{0\text{-}\pi}^{(2)} + H_{\rm int} + H_{\rm drive}^{(1)},
\end{align}
where $H_{\rm drive}^{(1)}$ represents two microwave drives on $n_{\theta 1}$, each with the form in \cref{eq:zp_H_drive}. 

\begin{table*}[!htbp]
\centering
\setlength{\tabcolsep}{3.5pt}
\renewcommand{\arraystretch}{1.2} 
\begin{tabular}{cccc cccc cc}
\hline \hline
Gate & $n_{\rm op}$ & Transitions & Associated $n_{\rm op}$ & $t_g$ & $\Omega_1$  & $\Omega_2$  & $\delta_1$ & $\delta_2$ & $F_{\rm optimize}$   \\ 
 &  &  &  & (ns) & ($2\pi\cdot$MHz) & ($2\pi\cdot$MHz) & ($2\pi\cdot$MHz) & ($2\pi\cdot$MHz) &   \\ 
\hline 
CNOT & $n_{\theta 1}$ & $[\ket{0,2} $-$ \ket{8,2}$, $\ket{2,2} $-$ \ket{8,2}]$ & $[0.037, 0.057]$ & 170.00 & 8.28 & 8.28 & -8.23 & -7.89 & 99.229\%  \\ 
\hline \hline
\end{tabular}
\caption{
The parameters of the CNOT gate corresponding to the red star in \cref{fig:cnot_population}. $A_i/2\pi$ is limited to be less than $100$ MHz. 
The effective Rabi rate is given by $\Omega_{i}=A_i(e^2-1)n_{\rm op}^{i}$ for $i\,=\,1,2$.
}\label{tab:parameter_cnot}
\end{table*}

\subsection{Transition selection and gate principle}\label{sec_sub:cnot_transition}

To identify a suitable Raman pathway, we examine the matrix elements of $n_{\theta 1}$ in \cref{fig:cnot_ntheta}(b). The rows correspond to computational states and the columns to low-energy non-computational states. A column containing two appreciable matrix elements identifies a candidate intermediate state for an effective $\Lambda$ system. Desirable pathways combine large matrix elements, which enable faster gates, with sufficient spectral isolation to suppress off-resonant transitions.

\Cref{fig:cnot_ntheta}(c) shows the relevant matrix elements as a function of transition frequency. Although the pathway through $\ket{4,5}$ is the most spectrally isolated, numerical optimization gives better gate performance using $\ket{8,2}$. This indicates that spectral isolation alone does not determine the fidelity: the complete set of matrix elements, detunings, and off-resonant couplings must also be considered. We therefore use $\ket{8,2}$ as the intermediate state.\footnote{The single-qubit $n_\theta$ gate involves $\ket{7}$. In the coupled circuit, charging-energy renormalization produces an avoided crossing and exchanges the roles of $\ket{7}$ and $\ket{8}$, so that $\ket{8}$ plays the analogous role here. See \cref{sec:crossing}.}

The inter-qubit coupling spectrally separates the Raman pathways through $\ket{8,0}$ and $\ket{8,2}$, allowing the latter to be driven selectively. Two tones couple $\ket{0,2}\leftrightarrow\ket{8,2}$ and $\ket{2,2}\leftrightarrow\ket{8,2}$, as shown in \cref{fig:cnot_ntheta}(d). The first qubit is therefore flipped only when the second qubit is in $\ket{2}$, implementing a CNOT with qubit 2 as the control and qubit 1 as the target. Residual phases are removed using single-qubit $Z$ rotations, as shown in \cref{sec:cnot_z_correction}~\cite{nesterov2022cnot}.

Other choices of Raman pathway produce different two-qubit gates. Coupling $\ket{0,0}$ and $\ket{2,2}$ gives a bSWAP gate, while coupling $\ket{0,2}$ and $\ket{2,0}$ gives an iSWAP gate.

\subsection{Gate operation and noiseless fidelity}\label{sec_sub:cnot_fidelity}
\begin{figure}[!htbp] 
    \includegraphics[width=1\columnwidth]{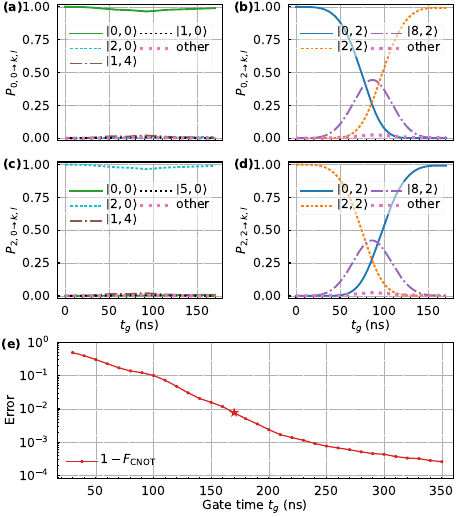}
\centering
\caption {Noiseless time evolution of the populations of various states during gate operation.  In subfigures (a)--(d),
the symbol $P_{i,j \rightarrow k,l }$ denotes the population transfer that starts at the state $\ket{i,j}$  and ends in the state $\ket{k,l}$.
The ``other'' state refers to the total population of all states not explicitly shown within the lowest 1000 dressed states.
(e) Average gate error ($1-F$) versus $t_g$ in the absence of noise. Each point corresponds to optimized pulse parameters for the given $t_g$. Star markers indicate the population transfer shown in (a)--(d).
}\label{fig:cnot_population}
\end{figure}

Panels (a)--(d) of \cref{fig:cnot_population} show the noiseless time evolution of various two-qubit state populations during the CNOT. The drive pulses are optimized in the same way as in the single-qubit case, with details given in \cref{sec:2Qgate_simulation}.  

The optimized gate has duration $t_{\rm g}=170$ ns and fidelity $F \sim 99\%$. As expected, only cases where the control qubit is in $\ket{2}$ show population transfer at the final time, as seen in \cref{fig:cnot_population}(b) and (d). The intermediate state $\ket{8,2}$ is substantially populated, reaching $>40\%$.

From \cref{fig:cnot_population}(a) and (c), we identify $\ket{8,0}$ as the dominant leakage state, from which we infer that the main leakage transitions are $\ket{0,0}$-$\ket{8,0}$ and $\ket{2,0}$-$\ket{8,0}$. These are plotted in green and red in \cref{fig:cnot_ntheta}(c) and (d). 
The leakage during the CNOT gate is small, which can be deduced from the population in unintentionally populated states (``other''). The total population in these states is approximately $0.66\%$, with no single state's contribution exceeding $0.15\%$. This indicates only minimal leakage during the gate.

Finally, \cref{fig:cnot_population}(e) shows the average gate error ($1-F$) versus $t_g$ in the absence of noise. Compared to the CZ gate discussed in the main text, this gate is much slower. Similar to the single-qubit lambda gates, it also exhibits some leakage. In principle, STIRAP could protect the gate, but competing factors such as decoherence arise. Ultimately, the slow operation time is the main limitation. For this reason we do not study decoherence for the CNOT gate.

Now we compare the CNOT gate with the CZ from the main text. The CZ gate achieves both shorter duration and higher fidelity. This advantage arises for two reasons: (i) CNOT requires driving two transitions in a $\Lambda$ system, while CZ uses direct transitions; and (ii) the relevant CZ transition has a charge matrix element an order of magnitude larger, with $\bra{2,0}n_{\theta 2}\ket{5,0}=0.537$ compared to $\bra{0,2}n_{\theta 1}\ket{8,2}=0.037$ and $\bra{2,2}n_{\theta 1}\ket{8,2}=0.057$ for CNOT (see \cref{tab:parameter_cz} and \cref{tab:parameter_cnot}).

\subsection{Single-qubit \texorpdfstring{$Z$}{Z} corrections for the CNOT gate}\label{sec:cnot_z_correction}
 For a bipartite Hilbert space $\mathcal{H}_1\otimes \mathcal{H}_2$, a controlled-NOT with control on subsystem $i$ and target on $j$ is denoted $\mathrm{CNOT}_{ij}$. 
The ideal $\mathrm{CNOT}_{21}$ is 
\begin{equation}
 \mathrm{CNOT}_{21}=   \begin{bmatrix}
        1 & 0 & 0 & 0\\
        0 & 0 & 0 & 1\\
        0 & 0 & 1 & 0\\
        0 & 1 & 0 & 0\\
    \end{bmatrix}.
\end{equation}
Our numerics yield a candidate gate $U_{\rm can}$ by projecting the simulated propagator (from the master equation) into the logical space. We parameterize its principal matrix elements as $z_{ij}= |z_{ij}|e^{i\phi_{ij}}$ and denote the small error terms collectively by $y$. The resulting form is
\begin{equation}
  {U}_{\rm can} =
 \begin{bmatrix}
  |z_{00}|e^{i\phi_{00}} & y & y & y \\
  y &  y & y & |z_{02}|e^{i\phi_{02}} \\
  y & y & |z_{20}|e^{i\phi_{20}} & y \\
  y & |z_{22}|e^{i\phi_{22}} & y & y 
 \end{bmatrix}.
 \end{equation}
The elements $|z_{ij}|$ are equal to 1 in the ideal operator, and the phases $\phi_{ij}$ capture the phase accumulated in time evolution that must be corrected.

To facilitate a comparison between this operator and the ideal CNOT, we apply additional single-qubit $Z$ rotations $R_z(\phi)=\exp[-i \phi Z/2]=\mathrm{diag} [e^{-i\phi/2},e^{i\phi/2}]$ both before and after the gate,
\begin{equation}
{U_{\rm f}}
=\! e^{-i\phi_{\rm gb}}
[R_z(\phi_2) \otimes R_z(\phi_1) ]
\cdot {U}_{\rm can} \cdot [ R_z(\phi_3)\otimes I ] \, .
\end{equation}
Here, $\phi_{\mathrm{gb}} =\phi_{00} - (\phi_1 +\phi_2 +\phi_3 )/2 $ is a global phase. To cancel out all phases of the key matrix elements, we solve linear equations to find the $R_z(\phi_i)$ rotation angles
\begin{equation}
\begin{split}
  \phi_1 &= \frac 12 \left(  \phi_{00} - \phi_{02} + \phi_{20} - \phi_{22}\right)\,,\\
  \phi_2 &=   \frac 12 \left(  \phi_{00} + \phi_{02} - \phi_{20} - \phi_{22}\right)\,,\\
  \phi_3 &= \frac 12 \left( \phi_{00} - \phi_{02} - \phi_{20} + \phi_{22}\right)\, \,.
  \end{split}
\end{equation}
 After using those angles, we arrive at
\begin{equation}
  {U}_{\rm f} =
 \begin{bmatrix}
  |z_{00}| & y' & y' & y' \\
  y' &  y' & y' & |z_{02}| \\
  y' & y' & |z_{20}| & y' \\
  y' & |z_{22}| & y' & y' 
 \end{bmatrix}\,,
 \end{equation}
  where the entries denoted by $y'$ are complex numbers and satisfy $\left|y'\right| \ll 1$ in the high-fidelity limit. We then compute the gate fidelity between $U_{\rm f}$ and the ideal $\mathrm{CNOT}_{21}$. A similar derivation for a $\mathrm{CNOT}_{12}$ is presented in Ref.~\cite{nesterov2022cnot}. 

\begin{figure}[!thbp] 
    \includegraphics[width=\columnwidth]{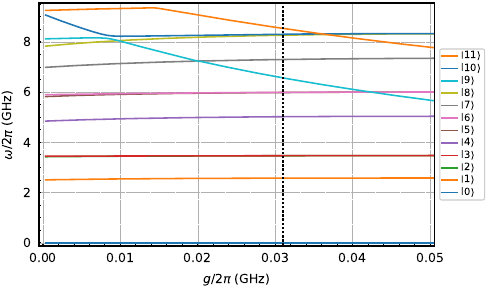}
\centering
\caption {
Energy spectrum of the first \zps qubit including coupling capacitances $C_c$ versus coupling rate $g$. The black dashed line is the operating point of the coupled qubits, $g/2\pi \approx 0.031$ GHz. Levels $\ket{7}$ and $\ket{8}$ cross near $g/2\pi \approx 0.02$ GHz.
}\label{fig:crossing}
\end{figure}

\subsection{Level Crossing in Coupled Circuits}\label{sec:crossing}
This section examines the level crossing between $\ket{7}$ and $\ket{8}$ as two \zps circuits are coupled. \Cref{fig:crossing} shows eigenvalues of the first \zps qubit versus coupling rate $g$, with $g=0$ for uncoupled systems. Eigenvalue shifts arise solely from the renormalized charging energies in \cref{eq:renormalized_ec}. The coupling rate depends on the capacitance $C_c$ (\cref{eq:coupling_rates_app}); larger $C_c$ yields stronger coupling. For our gates, we use $E_{C_c}/h=1.0$ GHz (black dashed line), giving $g/2\pi\approx0.031$ GHz.

There are two important level crossings at  $g/2\pi \approx 0.01$ GHz, between $\ket{8}$ and $\ket{9}$, and at $g/2\pi \approx 0.02$ GHz, between $\ket{7}$ and $\ket{8}$. These level crossings arise because the coupling capacitances renormalize the charging energies. Because $\ket{7}$ is used as an intermediate state in the $X$ gate, we expect $\ket{7,2}$ to be an important candidate state for two-qubit gates when there is no level crossing. However, due to the level crossing, $\ket{8,2}$ ends up being the intermediate state in the CNOT gate, while $\ket{7,2}$ has a nearly zero charge matrix element with respect to the logical states (not shown).  
We therefore expect the best CNOT performance near $g/2\pi\approx0.031$ GHz. This is the region where the intermediate state $\ket{8,2}$ is most isolated.

\begin{figure*}[!htbp]     
\includegraphics[width=0.99\textwidth]{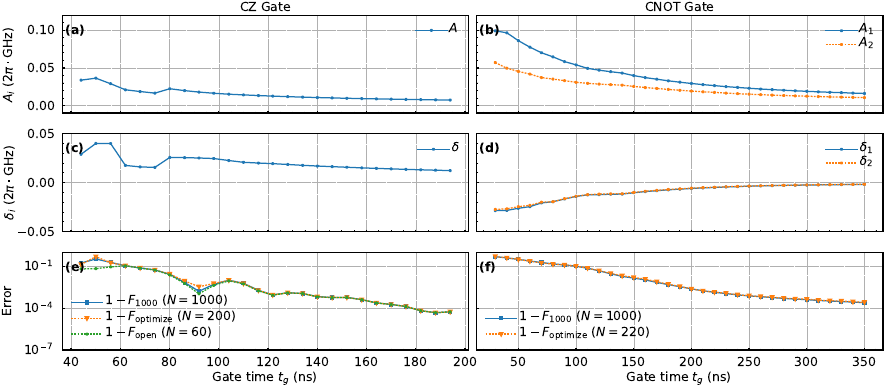}
\centering
\caption {
Optimized pulse parameters and gate infidelities for the CZ and CNOT gates. For each gate time $t_g$, panels (a,b) show the optimized drive amplitude $A$, panels (c,d) show the optimized detuning $\delta$, and panels (e,f) show the corresponding infidelities. For each fixed $t_g$, the Gaussian pulse is optimized in a truncated closed-system model of two coupled \zps qubits with $N_{\rm optimize}=200$ or $220$ selected levels, depending on the gate. The resulting fidelity is $F_{\rm optimize}$. The same optimized pulse is then tested in a closed-system model with $1000$ selected levels, giving $F_{1000}$,
and in a smaller ``selected'' $N$-level model, giving $F_{\rm open}$. The ``selected'' $N$-level model is used for open-system simulation including energy relaxation and pure dephasing. In this figure we only compare the closed-system fidelity for these three models. Note that we were unable to simulate the CNOT gate in an open-system due to the large model needed for convergence ($N_{\rm open}=240$).
The plotted infidelities are $1-F_{\rm optimize}$, $1-F_{1000}$, and $1-F_{\rm open}$.
}\label{fig:params_2q}
\end{figure*}

\section{Two-Qubit Gate Simulation Details}\label{sec:2Qgate_simulation}
The software used to generate the two-qubit gate simulation results is available in the repository \cite{zp2q_repo}. This appendix provides a high-level overview of the numerical techniques used to simulate CZ and CNOT gates, along with key implementation details. The goal is to help readers understand the core methods without requiring a deep dive into the source code.

\subsection{Noiseless simulation of two-qubit gates}
To simulate two-qubit gates, we first diagonalize the single-qubit Hamiltonians $H_{0\text{-}\pi}^{(1)}$ and $H_{0\text{-}\pi}^{(2)}$ to obtain eigenstates $\ket{p}^{(j)}$ and eigenvalues $E_p^{(j)}$, with $j = 1, 2$. We retain the lowest 300 eigenstates for each qubit. The eigenstates of the uncoupled system ($H_{\rm int} = 0$) are denoted $\ket{k}_0\otimes\ket{l}_0=\ket{k,l}_0$, where $k$ and $l$ label the eigenstates of the first and second qubits, respectively.

Next, we express the interaction term $H_{\rm int}$, e.g., $H_{\rm int}\approx  g_{\theta} n_{\theta_1}n_{\theta_2}$, in the uncoupled basis:
\begin{equation}
H_{\rm int} = \sum_{p,q,r,s=0}^{M_{\rm max}} 
c_{pqrs}\, \ket{p}_0\bra{q} \otimes \ket{r}_0\bra{s},
\end{equation}
where $c_{pqrs} = g_\theta \,_0\bra{p,r} (n_{\theta 1}\otimes n_{\theta 2} )\ket{q,s}_0$.
We then diagonalize the full coupled Hamiltonian $H_{\rm cp} = H_{0\text{-}\pi}^{(1)} + H_{0\text{-}\pi}^{(2)} + H_{\rm int}$ to obtain dressed eigenstates $\ket{p, q}$ and eigenvalues $E_{p,q}$. The resulting time-independent Hamiltonian $H_{\rm cp}$ is supplemented with a drive Hamiltonian and used to evolve the system under the Schr\"odinger equation:
\begin{equation}
i\frac{\partial V(t)}{\partial t} = \left( H_{\rm cp} + H_{\rm drive} \right) V(t).
\end{equation}
The evolution operator $V(t_g)$ at the final time $t_g$ is evaluated on the logical subspace spanned by $\{ \ket{0,0}, \ket{0,2}, \ket{2,0}, \ket{2,2} \}$ to form the $4 \times 4$ effective gate matrix $\bar V$, with elements
\begin{equation}
\bar V_{k,l;k',l'} = \bra{k,l} V(t_g) \ket{k',l'}.
\end{equation}
This operator is generally non-unitary due to leakage, although it will be nearly unitary for a high-fidelity operation.

\subsection{Noisy simulation of two-qubit gates} \label{app:subsec_2Q_noise_details_and_converge}
We use a master equation with local (to each \zps system) energy relaxation ($L^r_{ij}$) and dephasing ($L^\varphi_l$). The Lindblad operators for each \zps are constructed as 
\begin{equation}\label{eq:app_2Q_lindblad}
    \mathfrak{L}^{(1)} = L \otimes I\,\, {\rm and} \,\, \mathfrak{L}^{(2)} = I \otimes L\,, 
\end{equation}
where $L \in \{L^r_{ij},\, L^\varphi_l\}$ and we have suppressed the indices needed to label all the operators. Note that the operators in \cref{eq:app_2Q_lindblad} are defined in the bare basis, but our simulations use the dressed basis. These operators must be transformed into the dressed basis before solving the master equation.

Due to computational limitations, we cannot compute the noisy dynamics for systems with more than approximately 150 levels. As seen in \cref{fig:params_2q}, we identify a subset of states (60 for the CZ gate) that accurately represents the coherent dynamics of these operations. As in the single-qubit gate case, we evolve the system in these reduced subspaces to obtain the propagator $e^{\mathcal{L} t}$  that maps the initial density matrix $\rho(0)$ to $\rho(t)$. The resulting superoperator is projected into the computational subspace and corrected with single-qubit $Z$ rotations. Finally, we compute the  average channel fidelity with the target CZ, according to \cref{eq:channel_fidelity}.

 \section{Sensitivity Analysis of Circuit Parameters}\label{app:sensitivity}

\begin{table}[!htbp]
\centering
\setlength{\tabcolsep}{11pt}
\renewcommand{\arraystretch}{1.1}
\small
\begin{tabular}{ccc}
\hline \hline
\multirow{2}{*}{Parameter}
& \multicolumn{2}{c}{Sensitivity (decades)} \\
\cline{2-3}
& $t_g = 92~\mathrm{ns}$ & $t_g = 182~\mathrm{ns}$ \\ \hline
$E_{C}$                 & 0.445  & 0.244  \\
$E_{CJ}$                & 0.000  & 0.000 \\
$E_{J}$                 & 0.549  & 0.336  \\
$E_{L}$                 & 0.242  & 0.000  \\
$E_{Cc}$ (coupling)      & 0.000  & 0.075  \\
$E_{C0}$ (ground)        & 0.000  & 0.000  \\
$\phi_{\mathrm{ext}}$  & 0.010  & 0.002  \\
$n_{g}^\theta$                & -0.001 & -0.001 \\
\hline \hline
\end{tabular}
\caption{
Robustness of the optimized CZ gates against fabrication-induced parameter variations and operating-point offsets. Each circuit parameter of the second 0-$\pi$ qubit is independently varied by $\pm5\%$ around its nominal value while all other parameters are held fixed. For circuit parameters, we take 21 uniformly sampled points for the statistics.
The offset parameters $n_{g}^{\theta}$ and $\phi_{\mathrm{ext}}$ are sampled uniformly in log-space over
the ranges $[10^{-6},10^{-2}]$ and $[10^{-8},10^{-4}]$, respectively.
The sensitivity is the median change in coherent gate error, measured in decades relative to the nominal-point gate error.
Negative values indicate that the median gate error over the sweep is slightly smaller than the nominal-point error.
Both optimized gates are robust against most fabrication variations, while the shorter 92-ns gate exhibits increased sensitivity to variations in $E_{C}$, $E_{J}$, and $E_{L}$. 
}
\label{tab:cz_sensitivity}
\end{table}

In this appendix, we investigate the robustness of the optimized gates against circuit-parameter variations and operating-point offsets. Each circuit parameter of the second \zps qubit is sampled uniformly in a range $\pm5\%$ around its nominal value, while all other parameters are held fixed. Note that the offset parameters $n_{g}^{\theta}$ and $\phi_{\mathrm{ext}}$ are sampled uniformly in log-space over the ranges $[10^{-6},10^{-2}]$ and $[10^{-8},10^{-4}]$, respectively. The pulse is then re-optimized, and the corresponding coherent gate error is evaluated. We use 21 points to generate the statistics in \cref{tab:cz_sensitivity}.

We examine the two pulses highlighted in \cref{fig:2q_fidelity}, whose parameters are summarized in \cref{tab:parameter_cz}.  The shorter pulse, with $t_g\approx92$ ns, lies near a sharp minimum in the gate-error curve and is therefore expected to be sensitive to variations in gate time and other parameters. By contrast, the $182$-ns pulse is not in a minimum, suggesting greater robustness to such variations. 

The results are summarized in \cref{tab:cz_sensitivity}. The reported sensitivity is defined as the median value of $\log_{10}[(1-F)/(1-F_0)]$ over the parameter sweep, where $F_0$ is the coherent fidelity at the nominal parameter point; thus, one decade corresponds to a tenfold increase in coherent gate error. The 182-ns gate is generally insensitive to these variations, with all parameters producing a median increase in coherent gate error of less than one order of magnitude. By contrast, the faster 92-ns gate is more sensitive to variations in $E_{C}$, $E_{J}$, and $E_{L}$, while remaining comparatively robust against changes in $E_{CJ}$, $\phi_{\mathrm{ext}}$, $n_{g}^{\theta}$, the coupling energy $E_{Cc}$, and the ground-capacitance energy $E_{C0}$. These results demonstrate the trade-off between gate speed and robustness, with the longer gate exhibiting substantially improved tolerance to parameter variations.

\newpage 
\bibliography{zeropi}

\end{document}